# A comprehensive ab-initio insights into the pressure dependent mechanical, phonon, bonding, electronic, optical, and thermal properties of CsV$_3$Sb$_5$ Kagome compound

M. I. Naher[1,3,4], M. A. Ali[2,3], M. M. Hossain[2,3], M. M. Uddin[2,3], S. H. Naqib[3,4*]

[1]Energy Conversion and Storage Research Section, Industrial Physics Division, BCSIR Laboratories, Dhaka 1205

[2]Department of Physics, Chittagong University of Engineering and Technology (CUET), Chattogram-4349, Bangladesh

[3]Advanced Computational Materials Research Laboratory, Department of physics, Chittagong University of Engineering and Technology (CUET), Chattogram-4349, Bangladesh

[4*]Department of Physics, University of Rajshahi, Rajshahi 6205, Bangladesh
*Corresponding author; Email: salehnaqib@yahoo.com

**Abstract**
In this paper, we have presented a comprehensive study of the physical properties of kagome superconductor CsV$_3$Sb$_5$ using the density functional theory (DFT) methodology. The structural, mechanical, electronic properties (band structure, electronic energy density of states, Fermi surface, and charge density distribution), atomic bonding, hardness, thermodynamics, optical properties, and their pressure dependences have been investigated for the first time. The calculated ground state lattice parameters and volume are in excellent agreement with available experimental results. The estimated single-crystal elastic constants ensured the mechanical stability of the compound, whereas phonon spectra endorse dynamical instability at zero pressure. The electronic band structure, density of states, and optical properties confirmed the metallic feature. The Pugh ratio and Poisson's ratio of the compound under study revealed the softness/ductility. The hardness of CsV$_3$Sb$_5$, estimated from several formulae, is quite low while the machinability index predicted good machinability with excellent lubricating properties. The compound shows tendency towards structural instability at a pressure around 18 GPa. The optical constants have also been studied to correlate them with electronic properties (band structure) and predict the possible applications of this compound. Both the mechanical and optical properties show directional anisotropy. CsV$_3$Sb$_5$ is predicted to be an efficient absorber of ultraviolet radiation. The compound is also an efficient reflector of visible light.

## 1. Introduction

Layered Kagome-lattice transition metals are emerging as an exciting platform to explore the frustrating lattice geometry and quantum topology. The discovery of a new family of quasi-two-dimensional layered Kagome metals $A$V$_3$Sb$_5$ ($A$ = K, Cs, and Rb) has attracted intensive attention and has become a topic of extensive research due to many quantum phenomena and novel physical properties [1]. The highlighted physical phenomena are the non-trivial topological states [2,3], charge density waves (CDW) [4], strongly correlated electronic phases including charge order [3,5], anomalous Hall effect (AHE) [2,6-8], sizable correlation effects, and pressure-dependent superconductivity ($T_c$ = 0.9 ~ 2.7 K) [4,8-14]. Moreover, the vanadium (V) based kagome metals undergo CDW phase transitions with the critical temperatures $T_{CDW}$ ~ 78-103 K [1,2,15]. This kagome material family is found to possess superconducting behavior under pressure without any structural phase transition [16]. The Kagome lattices are subdivided into three categories based on their electronic ground states: metallic, semi-metallic, and insulating.

Here, we have selected the Kagome metal $CsV_3Sb_5$ for study because of its number of captivating properties and potential applications that have been discovered to date [2,4,12,13,17-19]. It is the heaviest metal among the layered vanadium antimonides. The superconductivity appears with a $T_c$ of 2.5 or 2.7 K [2,4], the highest in the $AV_3Sb_5$ family, and that can be increased up to 8 K at 2 GPa of applied pressure [12,13]. It shows anomalous Hall conductivity which reaches values larger than most ferromagnetic metals [2]. The $CsV_3Sb_5$ was reported to undergo a charge-density-wave (CDW) transition about $T_{CDW}$ ~ 94 K [2]. Pressure dependence on structural, transition temperature, total magnetic moment, and phonon spectra were also studied [17]. These disclosers of $CsV_3Sb_5$ sparked immense interest in exploring its bulk physical properties in further detail.

To the best of our knowledge, the following physical properties of the Kagome system $CsV_3Sb_5$, such as single crystal structure, superconducting transition temperature, CDW transition, electronic band structure, phonon dispersion Debye temperature, and reflectivity have been studied so far, either experimentally or theoretically [20-22]. Strikingly, a number of important physical properties and their pressure dependences, e.g., electronic (DOS, charge density distribution, Fermi surface), mechanical, hardness, acoustic, thermo-physical, bonding, and optical parameters of $CsV_3Sb_5$ have yet to be unveiled in depth. Therefore, correlating these unexplored physical properties with the existing features of $CsV_3Sb_5$ is vital to unlock the potential of this kagome material for possible applications. The bridging of this research gap is the prime motivation of our present investigation.

Therefore, in this paper, we have performed ab-initio calculations regarding aforementioned physical properties of $CsV_3Sb_5$ employing the first-principles calculations. Moreover, we have also considered the pressure effects on the studied physical properties of $CsV_3Sb_5$.

We have structured the rest of the manuscript out as follows: a brief description of the computational method employed in this study can be found in Section 2. A thorough discussion of the properties studied with their possible implications was presented in Section 3. The key features are summarized along with the conclusions pertinent to this study in Section 4.

## 2. Computational methodology

This work was carried out by employing the density functional theory (DFT) [23,24] in the CAmbridge Serial Total Energy Package (CASTEP) code [25]. The exchange-correlation energy was evaluated using the generalized gradient approximation (GGA) [26] incorporating the Perdew-Burke-Ernzerhof scheme for solids (PBEsol) [27] since the Perdew Burke-Ernzerhof (PBE) scheme [28] are reported to overestimate equilibrium volume and lattice parameters. The coulomb interaction between valence electrons (VE) and ion cores was modeled with the Vanderbilt-type ultra-soft pseudopotential [29]. The VE configuration for the elements is: Cs-$5s^2$ $5p^6$ $6s^1$, V-$3s^2$ $3p^6$ $3d^3$ $4s^2$, and Sb-$5s^2$ $5p^3$. The optimized (minimum total energy and internal forces) structure $CsV_3Sb_5$ was found by applying the Broyden–Fletcher–Goldfarb–Shanno minimization scheme [30]. The plane-wave cutoff energy for the compound was set at 300 eV. The Monkhorst-Pack scheme was used for the Brillouin zone (BZ) sampling with a mesh size of 11×11×6 [31]. The convergence parameters were set as follows: total energy within ($5\times10^{-6}$ eV/atom), maximum force within (0.01 eV/Å), maximum lattice point displacement within ($5\times10^{-4}$ Å), maximum ionic force within (0.01 eV/Å), maximum stress within (0.02 GPa), and Gaussian smearing width within (0.1 eV). The Fermi surface calculations demand denser $k$-point meshes, 23×23×12 used for $CsV_3Sb_5$. All the DFT calculations were performed at with the default temperature (0 K) and at different hydrostatic pressures in GPa.

The hexagonal crystal structure of $CsV_3Sb_5$ has six elastic constants ($C_{11}$, $C_{33}$, $C_{44}$, $C_{66}$, $C_{12}$, and $C_{13}$), that were obtained employing the stress-strain method [32]. All the macroscopic elastic moduli of $CsV_3Sb_5$, such as the bulk modulus ($B$), shear modulus ($G$), and Young's modulus ($E$), were/are evaluated from those $C_{ij}$ by using the Voigt-Reuss-Hill (VRH) approximation scheme [33,34].

The frequency-dependent optical spectra of the compound are derived from the well-known complex dielectric function, $\varepsilon(\omega) = \varepsilon_1(\omega) + i\varepsilon_2(\omega)$. The imaginary part $\varepsilon_2(\omega)$ is obtained from the following formula [25]:

$$\varepsilon_2(\omega) = \frac{2e^2\pi}{\Omega\varepsilon_0} \sum_{k,v,c} |\langle \Psi_k^c | \hat{u}.\vec{r} | \Psi_k^v \rangle|^2 \ \delta(E_k^c - E_k^v - E) \qquad (1)$$

In this formula, $\Omega$ represents the volume of the unit cell, $\omega$ is the (incident) photon frequency, $\varepsilon_0$ defines the dielectric constant of the free space, e is the charge of electron, ***u*** (*unit vector*) is defining the polarization of the incident electric field, ***r*** is the position vector, and $\Psi_k^c$ and $\Psi_k^v$ are the conduction and valence band wave functions at a given wave-vector $k$, respectively. The real part of the dielectric function, $\varepsilon_1(\omega)$, has been evaluated via the Kramers-Kronig transformation once $\varepsilon_2(\omega)$ is known. All the other optical parameters, such as complex refractive index $n(\omega)$, absorption coefficient $\alpha(\omega)$, energy loss-function $L(\omega)$, reflectivity $R(\omega)$, and optical conductivity $\sigma(\omega)$, can be deduced from the known dielectric function [35-37].

The projection of the plane-wave (PW) states onto a linear combination of atomic orbitals (LCAO) basis sets [38,39] is used to understand the bonding nature in $CsV_3Sb_5$, invoking Mulliken bond population analysis [40]. For the bond population analysis, we used the Mulliken density operator written on the atomic (or quasi-atomic) basis as follows:

$$P_{\mu\nu}^M(g) = \sum_{g'}\sum_{\nu'} P_{\mu\nu'}(g')S_{\nu'\nu}(g-g') = L^{-1}\sum_k e^{-ikg}(P_k S_k)_{\mu\nu'} \quad (2)$$

and the net charge on atom $A$ is defined as,

$$Q_A = Z_A - \sum_{\mu \in A} P_{\mu\mu}^M(0) \quad (3)$$

where, $Z_A$ represents the charge on the atomic core.

## 3. Results and discussion
### 3.1. Structural optimization

The crystal structure (unit cell) of $CsV_3Sb_5$ is displayed in Fig. 1, which belongs to the hexagonal system with the space group P6/mmm [No. 191] and has layered feature. The primitive unit cell contains nine atoms with one Cs, three V, and five Sb atoms. The most notable feature is that the kagome lattice/net formed by V atoms and V-Sb slab sandwiched by the Sb2 layers. There are two distinct Sb sublattices. The Sb2 layers form graphene-like networks of Sb and sandwich the Kagome layer. On the other hand, the Sb1 site is at the centers of the V hexagons. Bond distances for the V–V, V–Sb, Sb–Sb, and Cs-Cs nearest neighbors are given in Fig. 1. The Wyckoff positions of the atoms in the unit cell are adopted from Ref. [41]. The calculated structural parameters of optimized $CsV_3Sb_5$ crystal are listed in Table 1, which manifest a good agreement with the previous experimental and theoretical values. The ground state lattice structure of the compound is necessary for reliable prediction the physical properties of $CsV_3Sb_5$ [42]. The pressure dependent optimized lattice parameters are also given in Table 1.

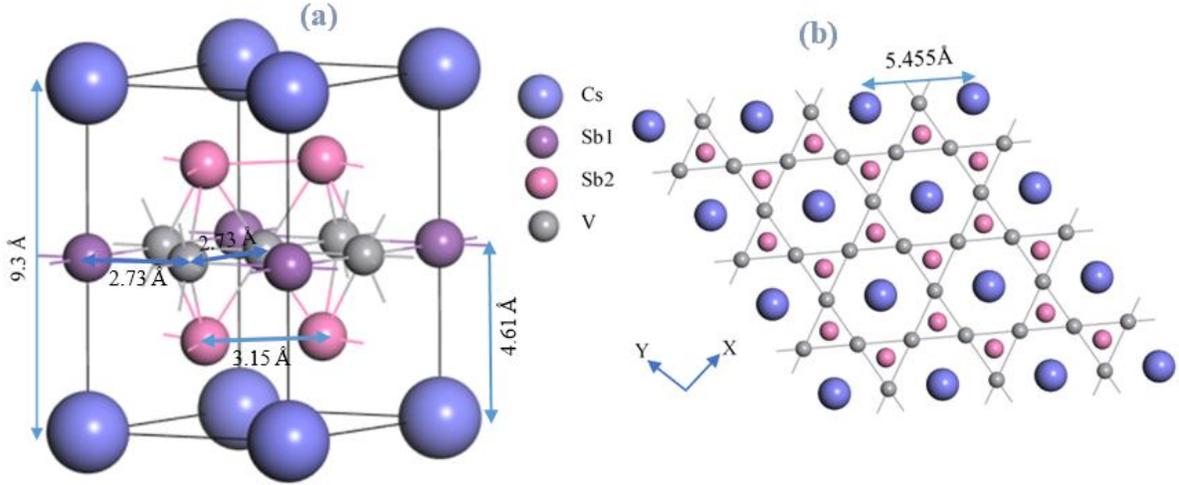

**Fig. 1.** (a) Schematic 3D Crystal structure of $CsV_3Sb_5$. (b) Top-view of the crystal structure (ab-plane), showing the kagome lattice of the $V_3Sb$ layer and the triangular lattice of the Cs layer. The gray lines show the kagome lattice of V. The crystallographic directions are shown.

**Table 1**
Calculated and experimental lattice constants (Å), equilibrium volume $V_0$ (Å$^3$), total number of atoms in the cell, and chemical unit $x$ of $CsV_3Sb_5$ under different hydrostatic pressures (GPa).

| P | a | C | c/a | $V_0$ | No. of atoms | x | Remarks |
|---|---|---|---|---|---|---|---|
| 0 | 5.455 | 9.230 | 1.692 | 237.83 | | | |
| 2 | 5.424 | 8.924 | 1.645 | 227.34 | | | |
| 4 | 5.408 | 8.424 | 1.558 | 213.34 | | | |
| 6 | 5.384 | 8.195 | 1.522 | 205.71 | | | |
| 8 | 5.357 | 8.102 | 1.512 | 201.36 | | | |
| 10 | 5.334 | 7.966 | 1.493 | 196.28 | 9 | 1 | This work |
| 12 | 5.315 | 7.870 | 1.481 | 192.54 | | | |
| 14 | 5.292 | 7.796 | 1.473 | 189.05 | | | |
| 16 | 5.270 | 7.723 | 1.465 | 185.72 | | | |
| 18 | 5.250 | 7.653 | 1.458 | 182.68 | | | |
| 20 | 5.231 | 7.593 | 1.452 | 179.92 | | | |
| - | 5.494 | 9.308 | - | - | - | - | [1,4][Expt.] |
| - | - | - | - | 236.60 | - | - | [16][Expt.] |
| - | 5.500 | 9.300 | - | - | - | - | [1][Expt.] |
| - | 5.400 | 9.000 | - | - | - | - | [43][Expt.] |
| 0 | 5. 450 | 9.297 | - | - | - | - | [44][Theo.] |
| - | 5.5097 | 9.306 | - | - | - | - | [45][Theo.] |

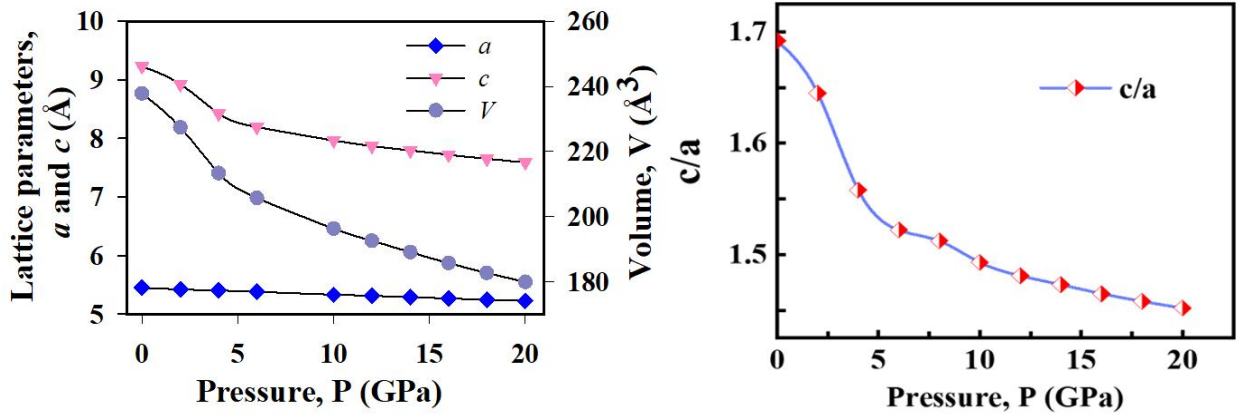

**Fig. 2.** (a) The unit cell parameters and equilibrium volume and (b) c/a ratio of CsV$_3$Sb$_5$ as a function of hydrostatic pressure.

### 3.2. Mechanical and elastic properties

The second-order elastic constants of materials are essential and indispensable parameters for characterizing their mechanical characteristics and drawing the limit to externally applied mechanical load for practical applications. Over the years, it has been played pivotal roles in many research fields, such as engineering, condensed matter physics, materials science, geophysics, and chemistry [46-50]. The stability, plasticity, stiffness, brittleness, ductility, chemical bonding, thermal characteristics, and anisotropy of material can all be calculated using its elastic constants. The independent elastic constants $C_{ij}$ for CsV$_3$Sb$_5$ at different pressures are given in Table 2. According to the Born-Huang criteria, mechanical stability conditions at equilibrium vary with crystal symmetries. The hexagonal CsV$_3$Sb$_5$ meets the Born stability criterion [51]: $C_{11} - |C_{12}| > 0$, $(C_{11} + C_{12})C_{33} - 2C_{13}^2 > 0$, $C_{44} > 0$, indicating that our compound is mechanically stable in the ground. Moreover, the mechanical stability criteria under pressure (P) for hexagonal crystal structure are as follows [51]:

$$\tilde{C}_{44} > 0, \tilde{C}_{11} > |\tilde{C}_{12}|, \tilde{C}_{33}(\tilde{C}_{11} + \tilde{C}_{12}) > 2\tilde{C}_{13}^2$$

where, $\tilde{C}_{ii} = C_{ii} - P (i = 1, 3)$ and $\tilde{C}_{1i} = C_{1i} + P (i = 2, 3)$
Based on these conditions, we can conclude that CsV$_3$Sb$_5$ shows mechanical instability at 18 GPa hydrostatic pressure since the value of $\tilde{C}_{44}$ is negative as seen in Table 2.

Each of the elastic constants has a different meaning; for example, $C_{11}$ and $C_{33}$ describe resistance to linear deformation and atomic bonding along [100] and [001], respectively. $C_{33} > C_{11}$ implies stronger atomic bonding along the *c*-axis than along the *a*-axis. Therefore, CsV$_3$Sb$_5$ is more compressible along [100]. Moreover, $C_{12}$ is associated with pure shear stress in the (100) plane along the [100] direction, and $C_{44}$ corresponds to the shear stress in the (010) plane along the [001] direction [52]. The constant $C_{44}$ represents shape change resistance with constant volume and provides a measure of the response to shear strain. The lower value of $C_{44}$ compared to $C_{66}$ suggests that the shear as being easier along the (100) plane than that along the (001) plane. Among the independent elastic stiffness constants, $C_{44}$ is closely related to the hardness and machinability index of solids. It is instructive to note that the axial, as well as shear deformations, are different for different directions, which is due to the differences in atomic arrangement and interaction among them.

**Table 2**

Calculated elastic constants ($C_{ij}$) (GPa) and machinability index ($\mu_M$) of CsV$_3$Sb$_5$ under different hydrostatic pressures (GPa).

| P | $C_{11}$ | $C_{33}$ | $C_{12}$ | $C_{44}$ | $C_{13}$ | $C_{66}$ | $\mu_M$ | Remarks |
|---|---|---|---|---|---|---|---|---|
| 0 | 127.983 | 151.433 | 53.094 | 29.872 | 71.056 | 37.445 | 2.936 | |
| 2 | 143.201 | 105.481 | 66.810 | 34.482 | 47.702 | 38.196 | 2.263 | |
| 4 | 176.883 | 115.529 | 83.282 | 37.983 | 59.410 | 46.801 | 2.471 | |
| 6 | 168.609 | 209.315 | 76.608 | 35.654 | 96.689 | 46.00 | 3.344 | |
| 8 | 203.981 | 116.278 | 94.302 | 42.815 | 47.527 | 54.84 | 2.206 | |
| 10 | 233.693 | 178.849 | 109.197 | 47.307 | 67.852 | 62.248 | 2.618 | This work |
| 12 | 218.081 | 140.607 | 95.060 | 28.128 | 49.933 | 61.511 | 3.668 | |
| 14 | 222.219 | 150.695 | 93.240 | 25.776 | 56.476 | 64.490 | 4.214 | |
| 16 | 237.074 | 130.146 | 106.904 | 9.816 | 49.010 | 65.085 | 10.726 | |
| 18 | 256.210 | 174.817 | 122.616 | -4.782 | 82.593 | 66.797 | -28.482 | |
| 20 | 313.306 | 198.929 | 172.863 | 26.503 | 109.392 | 70.222 | 6.448 | |

The machinability study of materials has been one of the important areas in the manufacturing industry. The material selection, processing, and machining operation can be planned efficiently to optimize the economic level if the machinability index of engineering materials is known. The cutting conditions (speed, force, depth, and energy), drilling rates, feed rate, tool wear, and machining time are considered as machinability assessment parameters. Machinability depends on many other external factors, such as work material, cutting tool geometry, cutting fluids, and temperature. The machinability of solids also reflects the lubricant and plasticity (resistance to plastic deformation) behaviors [53-55]. Plasticity increases with increasing machinability. The machinability index, $\mu_M$, of a material is defined as [56]:

$$\mu_M = \frac{B}{C_{44}} \tag{4}$$

Therefore, higher bulk modulus with lower shear resistance leads to better machinability and dry lubricity. Solids with good machinability exhibit excellent lubricating properties lower feed forces, lower friction values, and higher plastic strain values. The calculated value of $\mu_M$ is disclosed in Table 2. For comparison, experimental values for lubricants such as diamond, silver, and gold [57-58] are also given in Fig. 3. The high value of $\mu_M$ for the studied system in comparison with other materials [46-47] indicates excellent lubricating properties of CsV$_3$Sb$_5$, which may have potential in manufacturing industry. For pressures above 12 GPa, the machinability index of CsV$_3$Sb$_5$ exceeds that of gold (Au). Extremely high value of $\mu_M$ is found at 16 GPa. The negative value of the machinability index at 18 GPa is unrealistic and indicates that the structure is not relaxed and internal strain exists at this pressure.

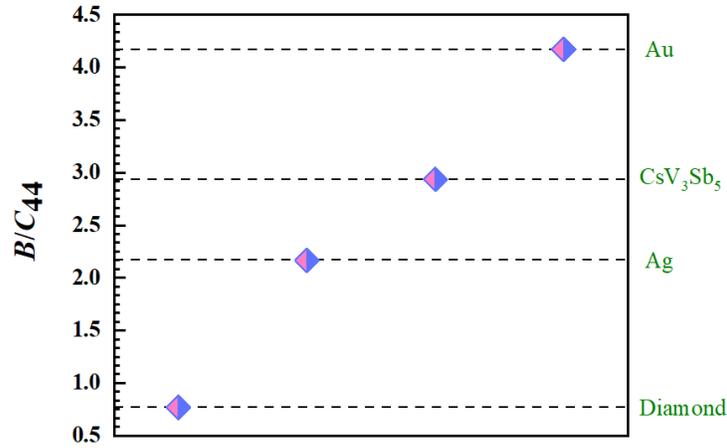

**Fig. 3.** Estimated machinability of CsV$_3$Sb$_5$ along with the experimental values for diamond, silver, and gold.

The elastic moduli ($B$, $G$, $E$, $v$), calculated from the second-order elastic constants, $C_{ij}$ using the Voigt-Reuss-Hill (VRH) formula [59-61], are listed in Table 3. The Reuss (R) and Voigt (V) approximations represent the upper and lower limits of $B$ and $G$, respectively. Moreover, Hill's (H) approximation estimates the bulk and shear moduli from the arithmetic mean of their Voigt and Reuss limits. The VRH approximated bulk, shear, and Young's moduli of the material were estimated using the following well-known formulae [62]:

$$B_V = \frac{[C_{11} + C_{22} + C_{33} + 2(C_{12} + C_{13} + C_{23})]}{9} \quad (5)$$

$$G_V = \frac{[C_{11} + C_{22} + C_{33} + 3(C_{44} + C_{55} + C_{66}) - (C_{12} + C_{13} + C_{23})]}{15} \quad (6)$$

$$B_H = \frac{B_V + B_R}{2}; \quad G_H = \frac{G_V + G_R}{2} \quad (7)$$

$$E_V = 9B_V G_V / (3B_V + G_V) \quad (8)$$

The obtained values are listed in Table 3. Henceforth, we drop the subscripts from the symbols denoting the polycrystalline VRH approximated elastic moduli. The fracture strength for all pure metals is proportional to their bulk modulus ($B$) and lattice parameter ($a$). The bulk modulus indicates the resistance to the volume change against external stress, indicating the average bond strength. The rigidity against plastic deformation or resistance to bonding angle is proportional to the shear modulus ($G$). The lower value of $G$ than $B$ predicts plasticity will dominate its mechanical strength (Table 3). The bulk modulus and the volume of the unit cell of a solid follow the following relationship: $B \sim V^k$ [42], with the exponent $k$ in the order of unity. The bulk modulus is proportional to the cohesive energy or the heat of sublimation [63] and represents, on average, the material's opposition to bond rupture [64]. The larger value of $G$ manifests the more pronounced directional bonding between atoms [65]. The ratio ($G/B$), proposed by Pugh is closely related to the brittleness and ductility of a material [64]. For brittle (ductile) material, the value of Pugh ratio is higher (lesser) than 0.57. The $G/B$ value of Kagome compound under study is 0.38, which confirms its ductility. Dislocation motion governs the strength and ductility of solids. Pugh also suggested that the strain ($\varepsilon$) at fracture can be measured from $\varepsilon \propto (B/G)^2$. Therefore, compounds with low Pugh ratio have high fracture strain. The fracture behavior of different materials is a commonly studied parameter in materials science. Tanaka *et al.* [66] proposed that $G/B$ represents the relative directionality of the bonding in materials. The Young's modulus ($E$) defines the stiffness of solids.

The Poisson's ratio ($\upsilon$) is another fundamental parameter characterizing the mechanical behavior of materials and is obtained from the following expression [52]:

$$\nu = \frac{(3B - 2G)}{2(3B + G)} \tag{9}$$

The Poisson's ratio measures compressibility, brittleness/ductility, and the bonding nature of solids. For all elastically stable solids, $v$ ranges between -1.0 and 0.5. A material's shape remains unchanged with any deformation for $v = 0.5$. Solids with $v > 0.26$ and $v < 0.26$ are expected to be ductile and brittle, respectively [67]. It provides more information about the characteristics of the bonding forces than any of the other elastic parameters [68]. The lower and upper limits of $v$ for central-force solids are 0.25 and 0.50, respectively [69-72]. It has been confirmed that borderline $v$ for covalent and ionic materials are 0.10 and 0.25, respectively [71]. It also correlates with Young's modulus ($E$) and compressibility ($1/B$). The estimated value of $v$ of $CsV_3Sb_5$ is 0.33 (Table 3). Therefore, we can predict that our compound is stable, ductile, and with a central inter-atomic force. The Poisson's ratio is also associated with the elastic/bonding anisotropy.

**Table 3**
The calculated isotropic bulk modulus $B$ (GPa), shear modulus $G$ (GPa), Young's modulus $E$ (GPa), Pugh's indicator $G/B$, Poisson's ratio $v$, and Cauchy pressures (GPa) of $CsV_3Sb_5$ under different hydrostatic pressures (GPa).

| P | B | | | G | | | $\frac{B_V}{B_R}$ | $\frac{G_V}{G_R}$ | G/B | E | N | $P_C^a$ | $P_C^b$ | Remarks |
|---|---|---|---|---|---|---|---|---|---|---|---|---|---|---|
| | $B_V$ | $B_R$ | $B_H$ | $G_V$ | $G_R$ | $G_H$ | | | | | | | | |
| 0 | 88.65 | 86.74 | 87.69 | 33.58 | 33.10 | 33.34 | 1.02 | 1.02 | 0.38 | 88.77 | 0.33 | 41.18 | 15.65 | This work |
| 2 | 79.59 | 76.47 | 78.03 | 36.74 | 36.37 | 36.55 | 1.04 | 1.01 | 0.47 | 94.85 | 0.30 | 13.22 | 28.61 | |

| | | | | | | | | | | | | | |
|---|---|---|---|---|---|---|---|---|---|---|---|---|---|
| 4 | 97.056 | 90.69 | 93.87 | 42.37 | 41.42 | 41.89 | 1.07 | 1.02 | 0.45 | 109.41 | 0.31 | 21.43 | 36.48 |
| 6 | 120.72 | 117.76 | 119.24 | 41.90 | 41.06 | 41.48 | 1.03 | 1.02 | 0.35 | 111.51 | 0.34 | 61.04 | 30.61 |
| 8 | 100.33 | 88.53 | 94.43 | 50.42 | 48.48 | 49.45 | 1.13 | 1.04 | 0.52 | 126.30 | 0.28 | 4.71 | 39.46 |
| 10 | 126.23 | 121.44 | 123.83 | 58.13 | 56.22 | 57.17 | 1.04 | 1.03 | 0.46 | 148.64 | 0.30 | 20.55 | 46.95 |
| 12 | 107.40 | 98.94 | 103.17 | 49.01 | 41.62 | 45.32 | 1.09 | 1.18 | 0.44 | 118.59 | 0.31 | 21.81 | 33.55 |
| 14 | 111.95 | 105.28 | 108.61 | 49.14 | 40.02 | 44.58 | 1.06 | 1.23 | 0.41 | 117.65 | 0.32 | 30.70 | 28.75 |
| 16 | 112.68 | 97.90 | 105.29 | 43.57 | 19.89 | 31.73 | 1.15 | 2.19 | 0.30 | 86.50 | 0.36 | 39.19 | 41.82 |
| 18 | 140.32 | 132.09 | 136.20 | 38.08 | -13.43 | 12.32 | 1.06 | -2.84 | 0.09 | 35.88 | 0.46 | 87.38 | 55.82 |
| 20 | 178.76 | 163.02 | 170.89 | 53.57 | 42.13 | 47.85 | 1.10 | 1.27 | 0.28 | 131.30 | 0.37 | 82.89 | 102.64 |
| 0 | - | - | 38.10 | - | - | - | - | - | - | - | - | - | - | [16]$^{Theo.}$ |

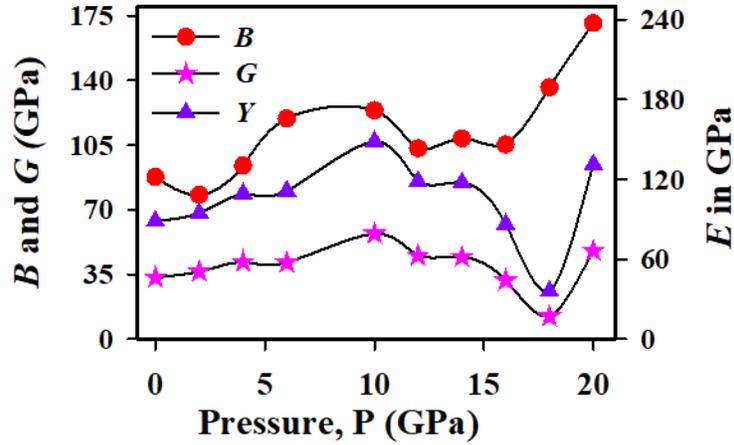

**Fig. 4.** Change in bulk, shear and Young's modulus of CsV$_3$Sb$_5$ with pressure.

Another fundamental mechanical parameter for material characterization is the Cauchy pressure, $P_C$. It reflects bonding nature at the atomic level. The expression of $P_C$ depends on the crystal symmetry and can be calculated for hexagonal symmetry as follows [73-75].

$$P_C^a = C_{13} - C_{44}, \; P_C^b = C_{12} - C_{66} \tag{10}$$

Here, $P_C^a$ and $P_C^b$ represent the Cauchy pressure for the (100) and (001) planes, respectively. The ductility of CsV$_3$Sb$_5$, predicted from the $G/B$ ratio, can be further strengthened by the Cauchy pressure [76]. The positive and negative Cauchy pressure suggests intrinsic ductility and brittleness of the material, respectively. The estimated Cauchy pressures of the compound for different planes are summarized in Table 3. It can also predict the angular characteristics of atomic bonding in solids. Pettifor's rule [73] says that the positive and negative $P_C$ of a solid reflect metallic non-directional and angular bonding, respectively. The more negative $P_C$, the more directional characteristic, the lesser bonding mobility, and more brittleness. For materials with zero $P_C$, simple pair-wise potentials can explain the bonding. The positive values of $P_C$ for the compound suggest ductility with metallic bonding as found from the Pugh's ratio and Poisson's ratio. The Cauchy pressure $P_C^a$ is more than twice of $P_C^b$, implying that the metallic character of the bonding in the (100) plane is more significant than that in the (001) plane.

The pressure dependent variations in the elastic moduli are shown in Fig. 4. It is observed that the variations are nonmonotonic with large dips in the values of $G$ and $E$ at around 18 GPa. This is an indication of structural instability which we will discuss in the subsequent sections.

Dislocations act as the carriers of plasticity in metals. Dislocation nucleation at crack tips has become a topic of substantial interest. The Peierls stress ($\sigma_P$) and crystal lattice resistance are lower in ductile materials [77], and these parameters quantify dislocation mobility. Peierls stress is defined as the minimum external stress required to move a stationary dislocation irreversibly at 0 $K$. The stress required to initiate such a movement is obtained from the following formula [78]:

$$\sigma_P = \frac{2G}{1-\nu} exp\left(-\frac{2\pi d}{b(1-\nu)}\right) \qquad (11)$$

where $b$ is the Burgers vector and $d$ is the interlayer distance between the glide planes. In case of dislocation dynamics, there are two types of activation energy: activation energy of diffusion ($Q^D$) and activation energy of dislocation glide ($U$). Both activation energies are related to the product of shear modulus ($G$) and lattice constant ($a$) as [79]:

$$Q^D, U \propto aG \qquad (12)$$

where, the quantity $aG$ is also defined as the measure of force constant. From Eqns. 11 and 12, one can extract the important parameters related to the dislocation dynamics at different pressures for $CsV_3Sb_5$ from the computed values of $G$ and $\nu$.

Hardness, the resistance to permanent deformation, is one of the foremost mechanical properties of materials, which gives quality assurance in industrial applications. The application areas include microelectronics, optoelectronics, and coatings for low-emission window glasses, tribological coatings, and heavy-duty products [80-85]. Hardness has persistent importance in understanding the relationship between hardness and other properties such as scratch resistance, surface durability as well chemical stability of a material. Cutting tools and wear-resistant coatings are made of hard materials. There is a correlation between the hardness and strength of materials [86]. Over the years, several semi-empirical formulae have been developed to correlate the hardness of crystalline materials with many properties, like the bond length, bond strength, charge density, iconicity, electronegativity, energy band gap, etc. [87-91]. The Vickers hardness ($H_V$) of $CsV_3Sb_5$ is estimated from the elastic moduli using the following macroscopic models [62,92-94]:

$$H_1 = 0.0963B \qquad (13)$$

$$H_2 = 0.0607E \tag{14}$$

$$H_3 = 0.1475G \tag{15}$$

$$H_4 = 0.0635E \tag{16}$$

$$H_5 = -2.899 + 0.1769G \tag{17}$$

$$H_6 = \frac{(1-2v)B}{6(1+v)} \tag{18}$$

$$H_7 = \frac{(1-2v)E}{6(1+v)} \tag{19}$$

$$H_8 = 2(k^2 G)^{0.585} - 3 \tag{20}$$

Here, $k$ (= $G/B$). Therefore, hardness correlates not only with bulk modulus but also with shear and Young's modulus. The crystal structure and energy band gap ($E_g$) of semiconductor determine the accuracy of the above methods in predicting hardness [87]. The estimated hardness values of the compound using those formulae are listed in Table 4. It has been found previously that for tetragonal, orthorhombic, and hexagonal crystals, $H_8$ is a reliable predictor of experimental hardness [62]. The computed hardness of CsV$_3$Sb$_5$ is quite low. The friction coefficient under oil lubrication decreases with increasing hardness. Based on hardness, solid lubricants can be divided into two broad categories: soft (hardness less than 10 GPa) and hard (hardness more than 10 GPa) [95-97]. Soft solid lubricating coatings use soft metals. Solid lubricants are critically important for numerous tribological systems' safe and smooth operations. However, both intrinsic and extrinsic properties of a material rule (macroscopic) hardness. This parameter roughly forecasts the hardness of crystals rather than the true hardness since many other factors other than elastic moduli control the hardness of solids. Hardness anisotropy [98] is also an interesting topic since the bonding strengths of different bonds along different crystallographic directions is generally different. As evident from Table 4, CsV$_3$Sb$_5$ is expected to be soft with low hardness values. From Table 4, we find significant indication of lattice softening at around 18 GPa, irrespective of the formalism used to determine the bulk hardness. This is another strong indication of structural and elastic instability.

**Table 4**
The calculated values of hardness (GPa) of CsV$_3$Sb$_5$ at different hydrostatic pressures (GPa).

| P | $H_1$ | $H_2$ | $H_3$ | $H_4$ | $H_5$ | $H_6$ | $H_7$ | $H_8$ | Remarks |
|---|---|---|---|---|---|---|---|---|---|
| 0 | 8.445 | 5.388 | 4.918 | 5.637 | 2.999 | 3.986 | 3.750 | 2.015 | |
| 2 | 7.514 | 5.757 | 5.391 | 6.023 | 3.567 | 3.547 | 4.937 | 3.787 | This work |
| 4 | 9.040 | 6.641 | 6.179 | 6.948 | 4.511 | 4.267 | 5.425 | 3.986 | |

| | | | | | | | | |
|---|---|---|---|---|---|---|---|---|
| 6 | 11.483 | 6.769 | 6.118 | 7.081 | 4.439 | 5.420 | 4.310 | 2.176 |
| 8 | 9.094 | 7.666 | 7.294 | 8.020 | 5.849 | 5.410 | 7.236 | 6.117 |
| 10 | 11.925 | 9.022 | 8.433 | 9.439 | 7.214 | 5.629 | 7.625 | 5.598 |
| 12 | 9.935 | 7.198 | 6.685 | 7.530 | 5.118 | 4.690 | 5.788 | 4.125 |
| 14 | 10.459 | 7.141 | 6.576 | 7.471 | 4.987 | 4.937 | 5.348 | 3.4973 |
| 16 | 10.139 | 5.251 | 4.680 | 5.493 | 2.714 | 4.786 | 2.896 | 0.695 |
| 18 | 13.116 | 2.178 | 1.817 | 2.278 | -0.720 | 6.191 | 0.361 | -2.481 |
| 20 | 16.457 | 7.970 | 7.058 | 8.338 | 5.566 | 7.768 | 4.085 | 1.334 |

## 3.3. *Elastic anisotropy*

Mechanical/elastic anisotropy is one of the key factors in engineering sciences, which concerns the performance of the material under external loading (stress/strain). Anisotropy indices quantify how the mechanical properties of a system have its direction dependency. Elastic anisotropy controls many physical processes, like the propagation of micro-cracks in materials, alignment or misalignment of quantum dots, phonon conductivity, defect mobility, and development of plastic deformation in crystals, and controls the mechanical durability of materials. Therefore, a comprehensive understanding of anisotropic mechanical behavior is crucial. Generally, crystal's covalent (directional) and metallic bonding dominate its anisotropic and isotropic nature, respectively [62,99]. The degree of anisotropy in atomic bonding along different crystallographic planes and directions can be explained by the shear anisotropy factors given below [65]:

For the {100} shear planes between the ⟨011⟩ and ⟨010⟩ directions is,

$$A_1 = \frac{4C_{44}}{C_{11} + C_{33} - 2C_{13}} \tag{21}$$

for the {010} shear plane between ⟨101⟩ and ⟨001⟩ directions is,

$$A_2 = \frac{4C_{55}}{C_{22} + C_{33} - 2C_{23}} \tag{22}$$

and for the {001} shear planes between ⟨110⟩ and ⟨010⟩ directions is,

$$A_3 = \frac{4C_{66}}{C_{11} + C_{22} - 2C_{12}} \tag{23}$$

The calculated shear anisotropic factors are summarized in Table 5. All three factors are equal to unity for materials with shear isotropy, and any deviation from it defines the degree of anisotropy. The computed values of $A_1$ (= $A_2$) and $A_3$ at 0 GPa are 0.87 and 1.0, respectively. The compound shows shear isotropy for the {001} planes. The estimated values predict that the compound is moderately anisotropic, but nonmonotonic changes with pressure are also noticeable.

The universal anisotropy index $A^U$, equivalent Zener anisotropy measure $A^{eq}$, anisotropy in shear $A^G$ (or $A^C$), and anisotropy in compressibility $A^B$ of crystals with any symmetry can be estimated from [100-102]:

$$A^U = 5\frac{G_V}{G_R} + \frac{B_V}{B_R} - 6 \geq 0 \tag{24}$$

$$A^{eq} = \left(1 + \frac{5}{12}A^U\right) + \sqrt{\left(1 + \frac{5}{12}A^U\right)^2 - 1} \tag{25}$$

$$A^G = \frac{G^V - G^R}{2G^H} \tag{26}$$

$$A^B = \frac{B_V - B_R}{B_V + B_R} \tag{27}$$

$A^U$ is one of the most widely used indices to quantify elastic anisotropy in solids due to its applicability for all possible crystal symmetries. It is the first anisotropy parameter among all other anisotropy indices, which accounts for both shear and bulk contributions. Besides, $G_V/G_R$ influences $A^U$ more than $B_V/B_R$. $A^U$ is zero for the case of locally isotropic single crystals, whereas the degree of deviation from zero exhibits anisotropy in materials. The calculated values of $A^U$ at different hydrostatic pressures are collected in Table 5.

For locally isotropic materials, $A^{eq}$ is equal to one. The calculated values of $A^{eq}$ at different pressures predict the anisotropy of the studied compound (See Table 5). $A^G$ and $A^B$ represent percentage anisotropy in shear and compressibility, respectively. For both parameters, 0 and 1 (100%) represent elastic isotropy and the maximum possible anisotropy, respectively. The lesser value of $A^G$ than $A^B$ (Table 5) indicates higher anisotropy in compressibility than in shear.

The universal log-Euclidean index ($A^L$) is considered as the most general definition of anisotropy. $A^L$ is defined as follows [48,103]:

$$A^L = \sqrt{\left[\ln\left(\frac{B^V}{B^R}\right)\right]^2 + 5\left[\ln\left(\frac{C_{44}^V}{C_{44}^R}\right)\right]^2} \tag{28}$$

Here, $C_{44}^V$ and $C_{44}^R$ refer to the Voigt and Reuss values of $C_{44}$, respectively.

$$C_{44}^R = \frac{5}{3}\left\{\frac{C_{44}(C_{11} - C_{12})}{3(C_{11} - C_{12}) + 4C_{44}}\right\} \tag{29}$$

And

$$C_{44}^V = C_{44}^R + \frac{3}{5}\left\{\frac{(C_{11} - C_{12} - 2C_{44})^2}{3(C_{11} - C_{12}) + 4C_{44}}\right\} \tag{30}$$

For extremely anisotropic crystallites, $A^L$ gives more absolute measures than $A^U$. It has been reported that $A^L$ ranges from 0 to 10.27, while 90% of anisotropic materials have $A^L$ less than 1 [103]. $A^L$ is zero for isotropic crystals and increases with the degree of anisotropy. The estimated values of $A^L$ at different hydrostatic pressures are tabulated in Table 5. The degree of anisotropy in CsV$_3$Sb$_5$ increases with pressure. It has been stated that materials with higher (lower) $A^L$ possess layered (non-layered) structural features [46,48,62,103]. The comparatively low value of $A^L$ implies that the studied compound exhibits a non-layered structure.

The linear compressibilities of CsV$_3$Sb$_5$ along $a$- and $c$-axis ($\beta_a$ and $\beta_c$) were evaluated from [104]:

$$\beta_a = \frac{C_{33} - C_{13}}{D} \quad \text{And} \quad \beta_c = \frac{C_{11} + C_{12} - 2C_{13}}{D} \tag{31}$$

with $D = (C_{11} + C_{12})C_{33} - 2(C_{13})^2$

The ratio between the coefficients ($\beta_c/\beta_a$) is unity for isotropic compressibility, while any disparity elucidates the anisotropy in the compressibility. The estimated values (see Table 5) exhibit considerable anisotropy and higher along the $a$-axis.

**Table 5**

Shear anisotropy factor ($A_1$, $A_2$, and $A_3$), the universal anisotropy index ($A^U$), equivalent Zener anisotropy measure ($A^{eq}$), anisotropy in shear ($A_G$), anisotropy in compressibility ($A_B$), universal log-Euclidean index ($A^L$), linear compressibility ($\beta_a$ and $\beta_c$) (TPa$^{-1}$), and their ratio $\beta_c/\beta_a$ for the CsV$_3$Sb$_5$ compound under different pressures.

| P (GPa) | $A_1 = A_2$ | $A_3$ | $A^U$ | $A^{eq}$ | $A_G$ | $A_B$ | $A^L$ | $\beta_a$ | $\beta_c$ | $\beta_c/\beta_a$ |
|---|---|---|---|---|---|---|---|---|---|---|
| 0 | 0.87 | 1 | 0.090 | 1.32 | 0.007 | 0.101 | 0.084 | 0.005 | 0.002 | 0.40 |
| 2 | 0.90 | 1 | 0.091 | 1.316 | 0.005 | 0.020 | 0.044 | 0.003 | 0.007 | 2.33 |
| 4 | 0.86 | 1 | 0.185 | 1.477 | 0.011 | 0.034 | 0.097 | 0.002 | 0.006 | 3.00 |
| 6 | 0.77 | 1 | 0.125 | 1.379 | 0.010 | 0.012 | 0.106 | 0.003 | 0.002 | 0.46 |
| 8 | 0.76 | 1 | 0.33 | 1.680 | 0.020 | 0.062 | 0.156 | 0.002 | 0.007 | 2.96 |
| 10 | 0.68 | 1 | 0.190 | 1.484 | 0.017 | 0.019 | 0.125 | 0.002 | 0.004 | 1.87 |
| 12 | 0.43 | 1 | 0.971 | 2.391 | 0.081 | 0.041 | 0.856 | 0.002 | 0.005 | 2.35 |
| 14 | 0.40 | 1 | 1.21 | 2.628 | 0.102 | 0.031 | 1.120 | 0.002 | 0.005 | 2.15 |
| 16 | 0.15 | 1 | 6.10 | 6.940 | 0.373 | 0.070 | 3.337 | 0.018 | 0.055 | 3.03 |
| 18 | -0.07 | 1 | -19.14 | -0.07 | 2.091 | 0.030 | 5.081 | 0.002 | 0.004 | 2.32 |
| 20 | 0.36 | 1 | 1.45 | 2.86 | 0.120 | 0.046 | 1.241 | 0.001 | 0.004 | 2.98 |

All the anisotropy indices disclosed in Table 5 show anomalous variation close to 16 – 18 GPa range. This large variation suggests that bonding characters in different directions and crystal planes are strongly affected by pressure in this particular range.

In addition, the elastic anisotropy arises from the linear (directional) bulk modulus and the shear anisotropy. Therefore, just studying shear anisotropy factors are not sufficient. The uniaxial bulk modulus of solids can either be determined from the pressure-dependent lattice parameter or through the single crystal elastic constants. The bulk modulus in the relaxed state and uniaxial bulk moduli along the *a*-, *b*-, and *c*-axis and anisotropies in the bulk modulus of $CsV_3Sb_5$ are from the following formulae [65]:

$$B_{relax} = \frac{\Lambda}{(1 + \alpha + \beta)^2} \quad ; \quad B_a = a\frac{dP}{da} = \frac{\Lambda}{1 + \alpha + \beta} \quad ; \quad B_b = a\frac{dP}{db} = \frac{B_a}{\alpha} \quad ; \quad B_c = c\frac{dP}{dc} = \frac{B_a}{\beta} \quad (32)$$

And

$$A_{B_a} = \frac{B_a}{B_b} = \alpha \quad ; \quad A_{B_c} = \frac{B_c}{B_b} = \frac{\alpha}{\beta} \quad (33)$$

with

$$\Lambda = C_{11} + 2C_{12}\alpha + C_{22}\alpha^2 + 2C_{13}\beta + C_{33}\beta^2 + 2C_{23}\alpha\beta$$

$$\alpha = \frac{(C_{11} - C_{12})(C_{33} - C_{13}) - (C_{23} - C_{13})(C_{11} - C_{13})}{(C_{33} - C_{13})(C_{22} - C_{12}) - (C_{13} - C_{23})(C_{12} - C_{23})}$$

And

$$\beta = \frac{(C_{22} - C_{12})(C_{11} - C_{13}) - (C_{11} - C_{12})(C_{23} - C_{12})}{(C_{22} - C_{12})(C_{33} - C_{13}) - (C_{12} - C_{23})(C_{13} - C_{23})}$$

where, $A_{B_a}$ and $A_{B_c}$ refer to bulk anisotropy along the *a*- and *c*-axis with respect to *b*-axis, respectively.

The calculated values are listed in Table 6. The bulk modulus anisotropy of the compound is in the order, $B_c > B_a (= B_b)$. Therefore, the compressibility of the crystal along the *c*-axis is the largest. In general, the uniaxial bulk modulus of a system is different and much larger than the isotropic bulk modulus. This is because the pressure in a state of uniaxial strain for a given crystal density differs from the pressure in a state of hydrostatic stress at the same density of the solid [65]. $A_{B_a} = A_{B_b} = 1$, and any departure from unity represent elastic isotropy and anisotropy, respectively.

**Table 6**

Anisotropies in bulk modulus along different axes of $CsV_3Sb_5$ at different hydrostatic pressure.

| P (GPa) | $B_{relax}$ (GPa) | $B_a$ (=$B_b$) (GPa) | $B_c$ (GPa) | $A_{B_a}$ | $A_{B_c}$ |
|---|---|---|---|---|---|
| 0 | 99.36 | 246.89 | 509.28 | 1 | 2.06 |
| 2 | 90.92 | 362.17 | 182.59 | 1 | 0.50 |
| 4 | 104.53 | 472.34 | 187.53 | 1 | 0.40 |
| 6 | 134.94 | 332.09 | 720.25 | 1 | 2.17 |
| 8 | 146.98 | 728.43 | 246.42 | 1 | 0.34 |
| 10 | 149.15 | 576.71 | 308.96 | 1 | 0.54 |
| 12 | 121.46 | 528.60 | 109.94 | 1 | 0.43 |
| 14 | 105.28 | 436.84 | 203.25 | 1 | 0.47 |
| 16 | 97.89 | 492.55 | 162.48 | 1 | 0.33 |
| 18 | 132.09 | 570.16 | 246.13 | 1 | 0.43 |
| 20 | 163.02 | 812.85 | 272.19 | 1 | 0.33 |

The 2D and 3D graphical visualization of the directional dependence of Young's modulus ($E$), linear compressibility ($\beta$), Shear modulus ($G$), and Poisson ratio ($v$) of the crystal $CsV_3Sb_5$ are also studied with the help of ELATE software [105]. The elastic stiffness matrixes required for this study are estimated using the CASTEP code. Uniform circular 2D and spherical 3D graphical representations are the manifestations of the isotropic nature of crystals. The greater the deviation from these ideal shapes, the higher the degree of anisotropy. The 2D projection on the $xy$-, $xz$-, and $yz$-planes, along with the 3D view of $E$, $\beta$, $G$, and $v$ for the compound, are depicted in Fig. 5. The curves with green and blue colors represent the minimum and the maximum points for the parameters, respectively. It is evident that all four parameters are isotropic in the $xy$-plane, while they are anisotropic in other planes. Both 2D and 3D graphical plots show the anisotropy order: $E < G < K < v$. ELATE also recounts a quantitative analysis reporting the minimum and maximum values of all parameters and their ratio, as listed in Table 7.

**Table 7**

The minimum and maximum values of Young's modulus (GPa), compressibility (TPa$^{-1}$), shear modulus (GPa), Poisson's ratio, and their ratios for $CsV_3Sb_5$ in the ground state.

| $E$ | | $A_E$ | $K$ | | $A_K$ | $G$ | | $A_G$ | $v$ | | $A_v$ |
|---|---|---|---|---|---|---|---|---|---|---|---|
| $E_{min}$ | $E_{max}$ | | $K_{min}$ | $K_{max}$ | | $G_{min}$ | $G_{max}$ | | $v_{min}$ | $v_{max}$ | |
| 85.348 | 95.667 | 1.121 | 2.249 | 4.640 | 2.063 | 29.872 | 37.445 | 1.254 | 0.209 | 0.432 | 2.070 |

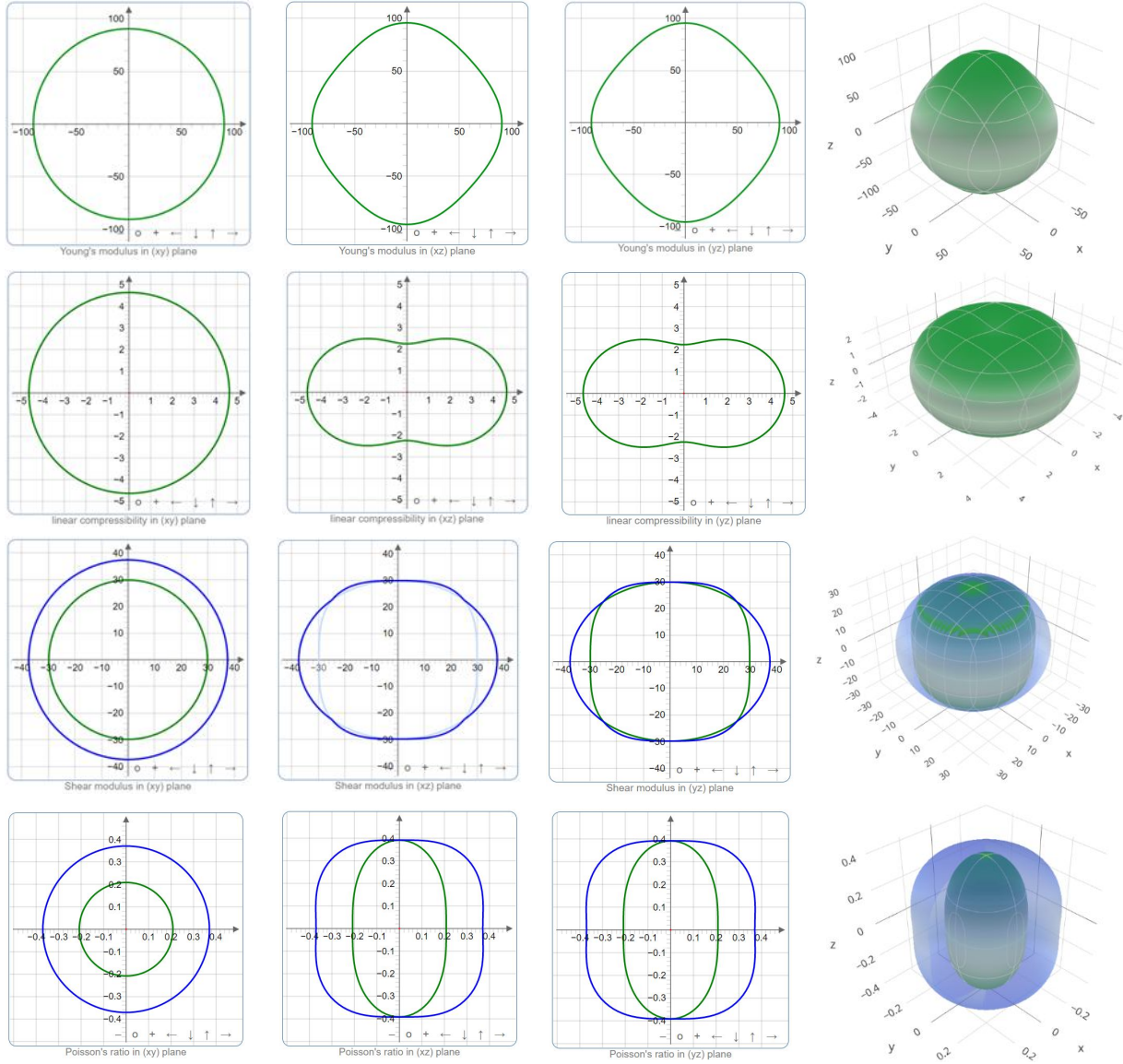

**Fig. 5.** Directional dependence of Young's modulus ($E$), compressibility ($\beta$), shear modulus ($G$), and Poisson's ratio ($v$) of $CsV_3Sb_5$.

### 3.4. Acoustic properties

The acoustic properties of interest in this study are: (i) the speed of sound and its anisotropy, (ii) acoustic impedance, and (iii) sound radiation coefficient. The second-order elastic constants contain information about how acoustic waves will propagate in a material. Sound velocity based techniques have been attracting considerable interest over the years to unveil the dynamic properties of materials in physics, materials science, seismology, geology, musical instrument design, and medical sciences [62]. Moreover, acoustic velocity and thermal conductivity ($k$) of solids are related [106]: $k = \frac{1}{3}C_v l v$. The transverse (shear), longitudinal, and average sound velocities through a crystalline material are defined by the following equations, respectively [107,108]:

$$v_t = \sqrt{\frac{G}{\rho}} \qquad v_l = \sqrt{\frac{B + 4G/3}{\rho}} \qquad \text{and} \qquad v_a = \left[\frac{1}{3}\left(\frac{2}{v_t^3} + \frac{1}{v_l^3}\right)\right]^{-1/3} \qquad (34)$$

where, $v_t$, $v_l$, and $v_a$ defines the transverse, longitudinal, and average sound velocities, respectively. $\rho$ refers to the mass-density of the crystal. The estimate acoustic velocities are summarized in Table 8. The variation in velocities with pressure is nonmonotonic. There have been ongoing attempts to classify materials by acoustic signals in developing humanoid robots' performances [109]. According to the sound velocity classification [109,110], our compound possesses low velocity. Such low sound velocity solids are useful as low thermal conductivity materials.

The acoustic impedance (*Z*), a widely used parameter in different branches of science, which defines the amount of acoustic energy that is transferred between two different materials. Like light waves, sound waves are also reflected/transmitted at the interface of two media with different acoustic properties. The speed of sound is higher in materials with higher *Z*. In terms of acoustic impedance, the ideally soft and hard surfaces are defined as $Z = 0$ and $Z \to \infty$, respectively. The acoustic impedance difference between the interfaces of the two media has a wide application in the musical industry, transducer design, acoustic sensors, aerospace industry, industrial factories, automobiles, medical ultrasound imaging, and in many underwater acoustic applications. The acoustic impedance of a solid is given by the expression [52]:

$$Z = \sqrt{\rho G} \qquad (35)$$

where, $\rho$ and *G* are the density and shear modulus of the medium. Denser and stiffer materials at a temperature have higher acoustic impedance. The calculated acoustic impedances for $CsV_3Sb_5$ at different pressures are displayed in Table 8. The degree of acoustic impedance mismatch between two media explains the amount of reflected and transmitted energy when a sound wave arrives at the interface. The acoustic impedance measurements of materials are essential to estimate the intensity of the sound reflected at an interface from the reflection coefficient (*R*) [111]:

$$R = \left(\frac{Z_1 - Z_2}{Z_1 + Z_2}\right)^2 \qquad (36)$$

Every vibrating material radiates acoustic energy and its intensity is considered an indicator of a material's suitability to attain perfect sound (pitch) as desired. The frequency of the pitch, radiated from a vibrating material, can be measured from $\sqrt{E/\rho}$ [112]. The acoustic radiation coefficient (*I*) assesses how much vibration of the body/specimen is damped due to acoustic radiation. The acoustic radiation coefficient of a material can be evaluated from the formula as follows [52,113]:

$$I = \frac{v}{\rho} = \sqrt{G/\rho^3} \text{ (m}^4/\text{kg.s)} \tag{37}$$

here, $v$ is the speed of sound. This parameter is crucial to choose proper materials for making high- and low-pitched musical instruments, for instance, violins, pianos, bells, tubas, and many more. The evaluated acoustic radiation coefficient is tabulated in Table 8. It is interesting to note that $CsV_3Sb_5$ has $I$ close to the Cu-Sn alloy. To give a general notion, relevant acoustic data for a widely used element, silver (Ag) at zero pressure, is presented in the third bracket (see Table 8). Acoustic converting efficiency (ACE) of a media is directly related to its acoustic radiation coefficient [114].

The Grüneisen parameter ($\gamma$) is directly related to phonon-phonon interaction strength and is generally a function of both volume and temperature. The acoustic ($\gamma_a$), elastic ($\gamma_e$), lattice ($\gamma_l$), thermodynamic ($\gamma_d$), and electronic ($\gamma_{el}$) are the five types of Grüneisen parameter. It is worth mentioning that for most metals, ionic and molecular crystals, the values of $\gamma_a$ and $\gamma_e$ coincide with the value of $\gamma_d$ [115]. On the contrary, for some materials, mostly rear-earth metals, a considerable difference occurs. The $\gamma_e$ value of $CsV_3Sb_5$ is evaluated from [116]:

$$\gamma_e = \frac{3(1+v)}{2(2-3v)} \tag{38}$$

This quantity exhibits various aspects of a material, such as thermal conductivity and expansion, the temperature dependence of elastic properties, and acoustic wave attenuation [62]. The effect of anharmonicity on $\gamma$ is considerable at higher temperatures (above Debye temperature). This parameter is useful for direct measure of "anharmonicity of the bonds" in a crystal [117]. For a crystal, the higher the $\gamma$, the higher anharmonicity, the higher thermal expansion, and the lower the phonon thermal conductivity. The value of $\gamma$ at 0 GPa obtained for $CsV_3Sb_5$ is 1.98 implying that the compound possesses significant anharmonicity. The pressure variation of the elastic Grüneisen parameter is nonmonotonic (Table 8).

**Table 8**
Density $\rho$ (g/cm$^3$), transverse velocity $v_t$ (ms$^{-1}$), longitudinal velocity $v_l$ (ms$^{-1}$), average elastic wave velocity $v_a$ (ms$^{-1}$), Grüneisen parameter $\gamma$, acoustic impedance $Z$ (Rayl), and acoustic radiation coefficient $I$ (m$^4$/kg.s) of $CsV_3Sb_5$ at different pressures (GPa).

| P | $\rho$ | $v_t$ | $v_l$ | $v_a$ | $\gamma$ | $Z$ (×10$^6$) | $I$ |
|---|---|---|---|---|---|---|---|
| 0 | 6.24 [10.4] | 2310.62 | 4600.10 | 2571.09 | 1.98 | 14.43 [27.9] | 0.37 [0.26] |
| 2 | 6.53 | 2365.36 | 4405.05 | 2641.87 | 1.77 | 15.45 | 0.36 |
| 4 | 6.96 | 2453.02 | 4637.57 | 2721.26 | 1.84 | 17.08 | 0.35 |
| 6 | 7.22 | 2396.99 | 4917.01 | 2689.94 | 2.08 | 17.31 | 0.33 |
| 8 | 7.38 | 2589.33 | 4662.91 | 2881.68 | 1.66 | 19.10 | 0.35 |
| 10 | 7.57 | 2761.47 | 5141.96 | 3058.79 | 1.77 | 20.80 | 0.36 |
| 12 | 7.71 | 2423.90 | 4605.29 | 2689.90 | 1.83 | 18.70 | 0.31 |
| 14 | 7.86 | 2382.15 | 4625.06 | 2647.39 | 1.90 | 18.71 | 0.30 |
| 16 | 8.00 | 1991.97 | 4296.21 | 2227.06 | 2.22 | 15.93 | 0.25 |
| 18 | 8.13 | 1231.01 | 4332.82 | 1402.84 | 3.53 | 10.01 | 0.15 |
| 20 | 8.25 | 2407.66 | 5332.15 | 2693.67 | 2.31 | 19.87 | 0.29 |

Large change in the values of sound velocities, Grüneisen parameter, acoustic impedance, and acoustic radiation coefficient close to 18 GPa is strong indicators of lattice instability.

## 3.5. Phonon dispersion

The calculation of the phonon dispersion spectra (PDS) and phonon density of states (PHDOS) of crystalline materials has become one of the fundamental research elements. It is instructive to check since it gives a lot of interesting information about the material regarding dynamic lattice stability/instability, phase transition and vibrational contribution to heat conduction, thermal expansion, superconducting $T_c$, Helmholtz free energy, and heat capacity [118-120]. The vibrational spectra results from the transition between quantized vibrational energy states of lattice. The lattice vibrations dominate in all temperature ranges and have pressure dependence. There is also a direct correlation between electron-phonon interaction function and the PHDOS. The PDS and PHDOS along the high symmetry directions of the Brillouin zone (BZ) have been calculated using density functional perturbation theory (DFPT) with the finite displacement method [121-122]. Fig. 6 shows the calculated phonon dispersion of $CsV_3Sb_5$ in the BZ in the ground state. No negative energy phonon branch exists. This implies dynamical stability of the Kagome compound at low pressure and temperature. A $2 \times 2$ CDW ground state due to Peierls instability is proposed in the kagome lattice [22]. We did not study pressure effect on the phonon modes since it has already been performed [17,18].

If a unit cell contains $N$ atoms, it will possess 3 acoustic modes and ($3N$-3) optical modes. The unit cell structure under consideration contains nine atoms, allowing 27 lattice vibrational modes, including 3 acoustic modes and 24 optical modes. The three branches of acoustic modes in the low-frequency region are one longitudinal and two transverse acoustic modes. The coherent vibrations of atoms in a lattice outside their equilibrium position are known as acoustic phonons. It is independent of the number of atoms the unit cell contains. Acoustic phonons contribute to sound and thermal transport in crystals. We have found that the low cut-off frequencies of the acoustic phonon modes that limit the phonon group velocity in different crystallographic directions of the Brillouin zone, resulting in a low lattice thermal conductivity [46,62]. The longitudinal acoustic vibrations in a material are connected with the electron scattering according to the Bloch-Grüneisen model [123]. Contrarily, the optical phonon is the out-of-phase oscillation of the atom in the lattice when an atom moves to the left and its neighbor to the right. Optical branches have non-zero frequencies at the $\Gamma$-point. Strong coupling between upper and lower optical branches is evident. The rapid flattening in the dispersion of the optical modes around 6 THz is observable for Sb atoms. The optical behavior of crystalline materials is strongly dominated by these optical branches. The directional dependency of phonon dispersion appears in our compound. The presence of phonon frequency gap between acoustic and optical phonons is closely related to the atomic mass ratio of the constituent atoms in the unit cell. The absence of any phononic band gap between the optical and acoustic branches might confirm the suitability of the compound for thermal transport without any break. The acoustic-optical phonon mode coupling can increase the scattering phase-space of phonons leading to reduced phonon lifetime and hence, a low $\kappa_{lat}$ or $k_{ph}$ [124]. The highest phonon modes ($E_{1u}$) happens around $\Gamma$- and $A$-points with the same value of ~8.5 THz.

The total PHDOS of the compound have been also calculated and displayed alongside the PDCs (Fig. 6). We have also computed the atomic PHDOS to estimate the contribution of each atom to different modes of vibrations (not shown). From the partial PHDOS, we found that acoustic dispersion curves are almost entirely due to vibration of the Cs atoms with small contributions from the V atoms. On the other hand, the optical modes correspond to the vibration of V, and Sb atoms. The optical branches around 6-8 THz mainly come from the vibration of the lighter Sb-atoms (See Fig. 1). The peaks in the total PHDOS are due to the flat parts of the PDCs as can be seen from Fig. 6.

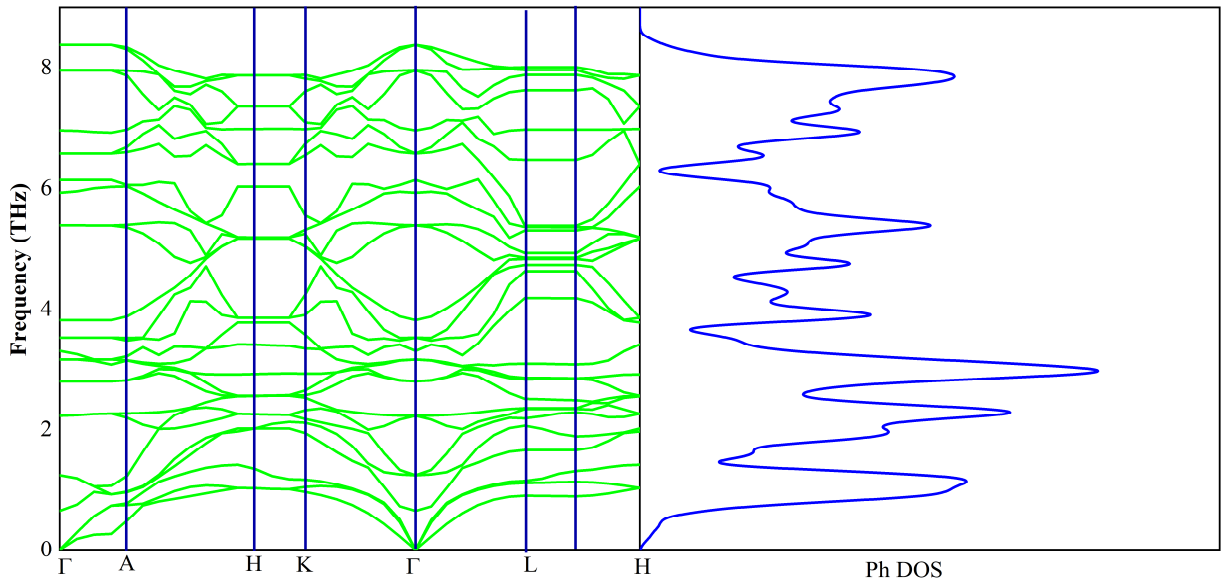

**Fig. 6.** The phonon dispersion spectra (PDS) and phonon density of states (PHDOS) of $CsV_3Sb_5$.

## 3.5. Thermophysical parameters

The parameters such as Debye temperature, melting temperature, thermal conductivity, thermal expansion coefficient, etc. are used to bring out the thermal behavior of a material, and study of those helps to predict potential applications of the system. We studied these properties under different hydrostatic pressures. All the calculated thermo-physical properties of $CsV_3Sb_5$ are summarized in Table 10.

### 3.5.1. Debye temperature

The Debye temperature of a material, one of the most widely studied physical parameters in solid-state physics, reflects the bonding strength between atoms and their vibrational energy. Besides, cohesive inter-atomic force strength in a crystal is directly related to the following properties: isothermal compression, indentation (microhardness), melting point, specific heat capacity, and thermal atomic movement. The stronger the cohesive forces, the higher resistance to compression, indentation, and thermal atomic motion. The higher the Debye temperature, the higher the phonon thermal conductivity of crystals. The Debye temperature of solids also distinguishes the classical and quantum-mechanical behavior of phonons. When the thermal energy in a system is higher than its Debye temperature, all the phonon modes get excited. The Debye energy represents the cutoff energy for the phonon modes. All modes of vibrations for a solid are expected to have the same energy (= $k_B T$) at temperatures higher than $\theta_D$. At the same time, the high-frequency modes are considered to be frozen for temperatures below $\theta_D$ [125]. The Debye temperature ($\Theta_D$) of $CsV_3Sb_5$ is calculated from the elastic constants, by using following formula [107]:

$$\Theta_D = \frac{h}{k_B}\left(\frac{3n}{4\pi V_0}\right)^{1/3} v_a \tag{39}$$

here, $h$ is Planck's constant, $k_B$ is the Boltzmann's constant, $V_0$ refers to the volume of unit cell, and $n$ defines the number of atoms in the cell. $\Theta_D$ is a function of temperature and decreases with increasing temperature due to the temperature dependence of elastic constants and sound velocity of the solid. Below and above $\Theta_D$, the acoustic and optical phonon modes are expected to be more prevalent in the lattice, respectively. Weak electron-phonon coupling is expected for all temperatures below $\Theta_D$. Since the electron mobility increases with temperature, the optical modes strongly interact with the electrons in solids than the acoustic modes. For temperature comparable to the Debye temperature, the phonon thermal resistivity and the electron-phonon coupling resistance increase as $T$ (due to the Umklapp phonon scattering) and as $\sqrt{T}$, respectively [126]. The phonon-phonon scattering rapidly increases as a function of temperature and vice versa. The calculated Debye temperature of the compound at 0 K and 0 GPa is 251.92 K (Table 10). Debye temperatures at different hydrostatic pressures are also listed in Table 10. In structurally stable solids, the Debye temperature increases with increasing pressure. Anomalous decrease in the Debye temperature of $CsV_3Sb_5$ at high pressures, particularly at 18 GPa is an indication of significant lattice softening of the compound.

### 3.5.2. Phonon thermal conductivity

In solids, both electrons and phonons can transport thermal energy. Electrons are dominant heat carriers in metals at low temperatures. At high temperatures, the lattice contribution becomes significant. The lattice (phonon) thermal conductivity of solids, $k_{ph}$, is one of the major thermophysical parameters with an enormous variety of technical applications, such as in the development of new thermoelectric materials, heat sinks, sensors, transducers and thermal barrier coatings [62,125,127]. Both low and high lattice thermal conductive materials have applications in different engineering branches. Finding low thermal conducting materials has got the focus in recent times to enhance thermoelectric performance, solid state refrigeration, and thermal barrier coating (TBC) systems. Conversely, high thermal conducting materials with minimum heat waste are crucial to improving the efficiency of heat remover in microelectronic and nanoelectronic devices. Compounds possess low thermal conductivity if they have heavy elements, large unit cells, and low melting points (i.e., low elastic moduli). All of these properties lead to low sound velocity. We have calculated $k_{ph}$ of CsV$_3$Sb$_5$ at 300 $K$, using the formula put forwarded by Slack, as follows [127]:

$$k_{ph} = A(\gamma) \frac{M_{av}\Theta_D^3 \delta}{\gamma^2 n^{2/3} T} \qquad (40)$$

here, $M_{av}$ is the average atomic mass per atom in the compound (kg/atom), $\Theta_D$ is the Debye temperature (K), $\delta$ refers to the cubic root of average atomic volume (m), $n$ represents the total number of atoms in the unit cell, $T$ defines the absolute temperature (K), and $\gamma$ is the Grüneisen parameter. $A(\gamma)$ is a constant (W-mol/kg/m$^2$/K$^3$) and can be calculated as follows [128]:

$$A(\gamma) = \frac{4.85628 \times 10^7}{2\left(1 - \frac{0.514}{\gamma} + \frac{0.228}{\gamma^2}\right)} \qquad (41)$$

The evaluated room temperature $k_{ph}$ of CsV$_3$Sb$_5$ at different pressures is enlisted in Table 10. The lower value of $k_{ph}$ indicates weaker covalent bonding. The low $k_{ph}$ is expected to be an indicator of the presence of soft phonon modes [129].

### 3.5.3. Melting temperature

The melting temperature ($T_m$) plays a vital role in identifying promising candidate materials in industrial applications for thermal management. For instance, low and high $T_m$ materials are good candidates for thermal interface material (TIM) and thermal barrier coatings, respectively. Solids at temperatures below $T_m$ are thermodynamically stable and can function continuously without oxidation, chemical change, and excessive distortion resulting in mechanical failure. The melting temperature increases with the bonding strength, higher cohesive energy, and lower coefficient of thermal expansion [52]. The $T_m$ the material was calculated using following expression [130]:

$$T_m = 354\ K + \left(\frac{4.5\ K}{GPa}\right)\left(\frac{2C_{11} + C_{33}}{3}\right) \pm 300\ K \tag{42}$$

The estimated value of $T_m$ for CaV$_3$Sb$_5$ in the ground state is 965.10 K (Table 10). Both $k_{ph}$ and $T_m$ of a crystal are associated with the bonding strength. It is well known that the melting temperature of a substance has pressure dependence. Studying melting points of materials under extreme conditions is one of the interesting branches of physics because of its importance in shock physics, planetary science, astrophysics, geophysics, nuclear physics, and materials science. Therefore, the pressure dependence of the melting temperature of our compound is also studied. It is perceptible that the melting temperature for CaV$_3$Sb$_5$ increases with the increase in pressure (Table 10), although the variation is nonmonotonic.

### 3.5.4. Thermal expansion

The thermal expansion coefficient (TEC) is an intrinsic thermal property of materials associated with the anharmonic lattice vibrations. It quantifies many features of a material, such as thermal conductivity, specific heat, entropy, and isothermal compressibility. Low thermal expansion materials are of great interest in the ceramic industry for their widespread use in high anti-thermal shock applications, electronic devices, heat-engine components, spintronics devices, etc [62]. The TEC of a material can be calculated from shear modulus, $G$ (in GPa) using the following expression [52]:

$$\alpha = \frac{1.6 \times 10^{-3}}{G} \tag{43}$$

The dimensionless quantity $\alpha\gamma T$ is in general a measure of the anharmonicity of a lattice [131]. According to the Mie-Grüneisen theory [117], $\alpha \propto C_V$. The calculated values of α at 300 K at different pressures are given in Table 10.

**Table 10**
The Debye temperature $\Theta_D$ (K), lattice thermal conductivity $k_{ph}$ (W/m-$K$) at 300 K, melting temperature $T_m$ (K), and thermal expansion coefficient $\alpha$ (K$^{-1}$) of CsV$_3$Sb$_5$ under different hydrostatic pressures (GPa).

| Compound | P | $\Theta_D$ | $k_{ph}$ | $T_m$ | $\alpha$ (10$^{-5}$) | Remarks |
|---|---|---|---|---|---|---|
| CsV$_3$Sb$_5$ | 0 | 251.92 | 2.81 | 965.10 | 4.80 | This work |
| | 2 | 262.77 | 4.04 | 941.82 | 4.38 | |
| | 4 | 276.37 | 4.38 | 1057.94 | 3.82 | |
| | 6 | 276.61 | 3.19 | 1173.80 | 3.86 | |
| | 8 | 298.44 | 6.53 | 1140.36 | 3.24 | |
| | 10 | 319.49 | 6.89 | 1323.35 | 2.80 | |

| | | | | |
|---|---|---|---|---|
| 12 | 248.87 | 3.03 | 1219.15 | 3.53 |
| 14 | 280.00 | 3.93 | 1246.70 | 3.59 |
| 16 | 227.83 | 1.50 | 1260.44 | 5.04 |
| 18 | 150.08 | 0.16 | 1384.86 | 12.99 |
| 20 | 289.63 | 2.75 | 1592.31 | 3.34 |

## 3.6. Electronic properties
### 3.6.1. Electronic band structure

The study of the electronic band structure of a material is crucial to explain many properties like the chemical bonding, electronic transport, superconductivity, optical response, and magnetic order, at a microscopic level. Bands around the Fermi level dominate electronic properties. Good band topologies and reasonable effective masses of charge carriers are prerequisite for modeling nanostructures and electronic devices. Studying the electronic band structure of materials is required in catalyst design [132], accelerated "battery materials" discovery [133], and superconducting materials development [134]. Therefore, we have studied the electronic band structure to understand band topology of the Kagome material. The electronic energy band structures $[E(k)]$ of $CsV_3Sb_5$, calculated with a smearing width of 0.1 eV, on a discrete $k$-mesh along the high symmetry directions in the first Brillouin zone at various hydrostatic pressures are shown in Fig. 7 (a-e). The total number of bands for the compound under consideration is 109. Strongly hybridized bands consisting of V $3d$ and Sb $5p$ orbitals spread over an energy range from -6 eV to 1.7eV. After 1.7 eV, the orbitals of Cs and Sb atoms dominate.

The metallic nature is observed from the noticeable overlap of conduction bands and valence bands at the Fermi level ($E_F$). The bands crossing the Fermi level (horizontal dotted line placed at zero energy) in the ground state are displayed in different colors with their corresponding band numbers 37 and 38, respectively. These two bands are degenerate at several points in the Brillouin zone, which is a result of high symmetry. The topological nontrivial electronic band structure, slightly overlapping bands (with multiple Dirac points) at certain points ($M$, $K$, $L$, and $H$ points) in the BZ, is also observed. Almost linear dispersive bands are observable in certain parts of the $E(k)$ diagram. For example, the quasi-linear electronic dispersions near the Fermi level at $L$- and $H$-points are found (shown with the brown circles). Such quasi-linear energy dispersion implies that the effective masses of the charge carriers with the associated wave vectors are significantly small, giving rise to a very high degree of mobility. The topological electronic band structures can be characterized by band inversion between the $\Gamma_6$ and $\Gamma_8$ energy levels at the $\Gamma$ symmetry point in the Brillouin zone [135]. The energy difference between $\Gamma_6$ and $\Gamma_8$, $\Delta E = \Gamma_6 - \Gamma_8$, is defined as the band inversion strength (BIS). BIS would be positive and negative for topologically nontrivial and trivial systems, respectively. Higher the BIS value greater is the phase stability. The bands crossing at the Fermi level increases to 3 at 15 GPa. With pressure the band gap increases around Γ-A. Moreover, the higher energy dispersion along

*c*-axis shows the anisotropy in electrical conductivity. It is noticeable from band structures that the electrical conductivity along the *c*-direction should be higher than that in the *ab*-plane. Both the electron- and hole-like features are observable in different sections in the BZ.

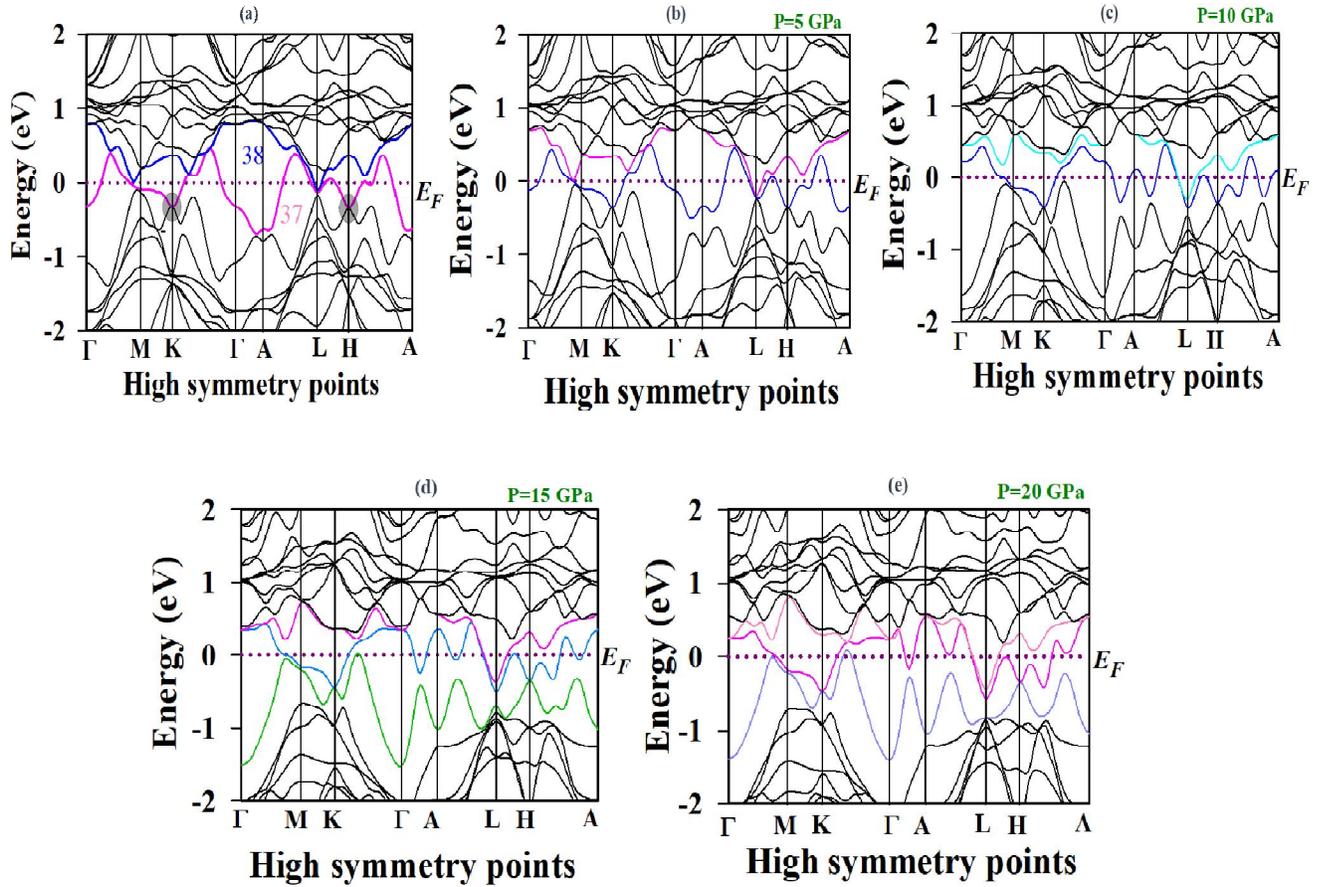

**Fig. 7.** The electronic band structures of $CsV_3Sb_5$ along the high symmetry directions in the BZ at (a) 0 GPa, (b) 5 GPa, (c) 10 GPa, (d) 15 GPa, and (e) 20 GPa.

### 3.6.2. Density of states (DOS)

It is instructive to study the total and partial electronic density of states (EDOS) of a material for better understanding of the bonding nature between atoms and the contribution of different atoms/orbitals to conductivity (thermal, optical, and electrical) and other electronic transport properties. Moreover, it also determines various vital physical quantities, like electronic contribution to the heat capacity in metals and spin paramagnetic susceptibility, which are directly related to the electronic density of states at the Fermi level, $N(E_F)$ [136,137]. Figs. 8(a-e) show the total and partial EDOS of $CsV_3Sb_5$ at different hydrostatic pressures. The vertical broken line at zero energy represents the Fermi level. The finite total EDOS at the Fermi level shows the metallic nature of the compound. From the EDOS figure, we can see that the low energy valence bands (not shown here) in the range of -6 eV to -2 eV are formed mainly from the Sb $5p$ electronic states. The bands around the Fermi level arise mainly from the strong hybridization between the V $3d$ and Sb $5p$ electronic states, where Cs atoms are also partly involved. Such hybridization near the Fermi energy is a useful indicator of the formation of strong covalent bondings [62]. Since the $3d$ states of V atoms heavily dominate the EDOS close to the Fermi level, we predict that these $3d$ electrons will dominate the electronic properties more than any other atomic orbitals. On the other hand, the total EDOS in the conduction band around 2 eV to 8 eV comes from the hybridization of the Cs $5p$ and $6s$ states, and the Sb $5p$ and $5s$ states. The atom resolved partial density of states (PDOS) of the V atom shows the most significant changes with pressure. The calculated total EDOS of $CsV_3Sb_5$ at $E_F$ is 5.44 states per eV per unit cell in the ground state which shows good agreement with the literature [22]. The Fermi level tends to shift to the right of the pseudogap with increasing pressure. This is an indication that the electronic stability of $CsV_3Sb_5$ weakens with increasing pressure.

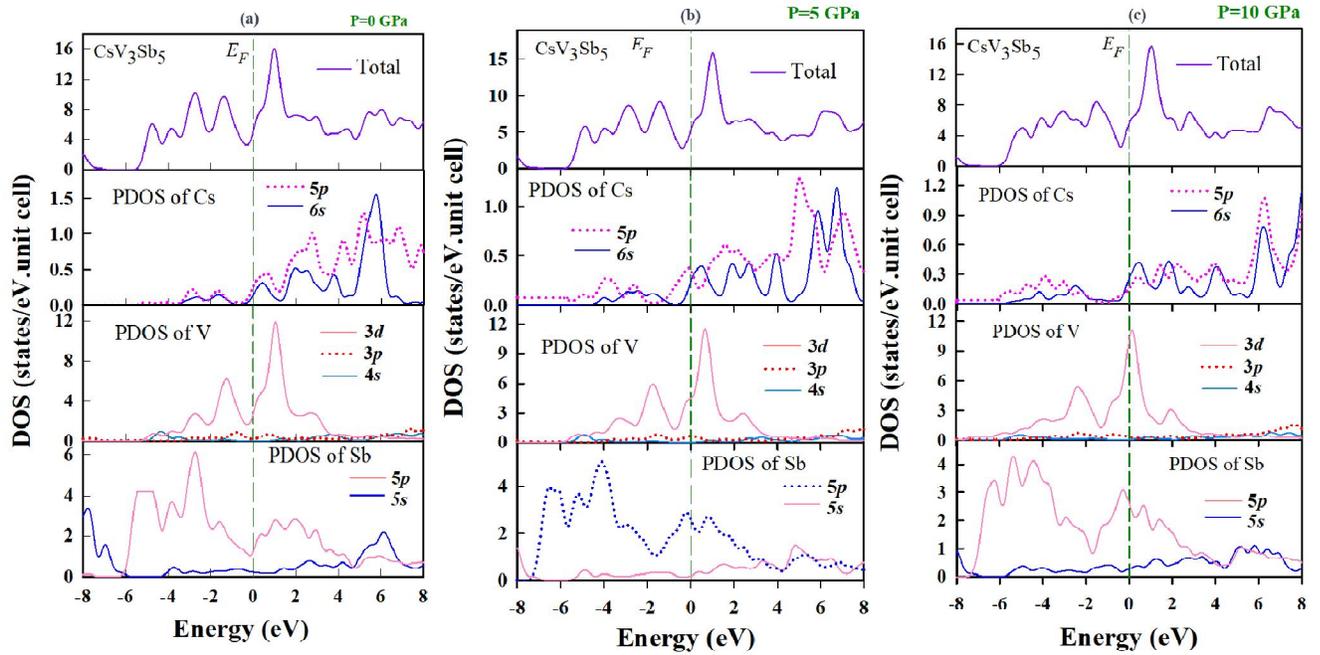

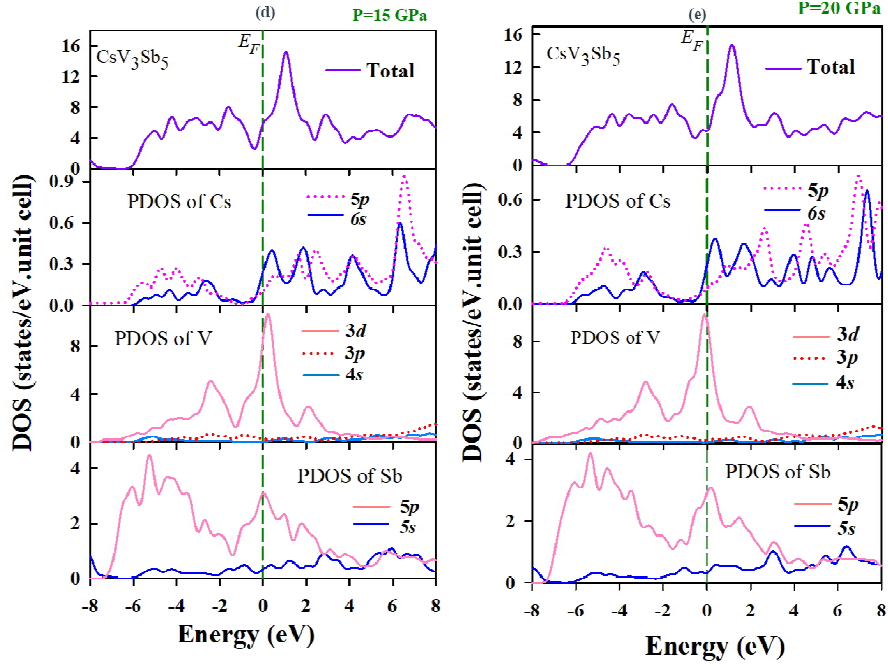

**Fig. 8.** Total and partial electronic density of states (PDOS and TDOS) of CsV$_3$Sb$_5$ as a function of energy at (a) 0 GPa, (b) 5 GPa, (c) 10 GPa, (d) 15 GPa, and (e) 20 GPa. The Fermi level is placed at zero energy.

The electron-electron interaction (EEI) due to Coulomb force, also known as the repulsive Coulomb pseudopotential plays an important role in determining electronic correlations in a system. It helps us to get an idea about the strength of the direct Coulomb repulsion between electrons. The Coulomb pseudopotential, $\mu^*$, can be gauged from the following expression [138]:

$$\mu^* = \frac{0.26 N(E_F)}{1 + N(E_F)} \qquad (44)$$

The estimated Coulomb pseudopotential for CsV$_3$Sb$_5$ in the ground state is quite high, 0.219 implying strong electronic correlations. The repulsive Coulomb pseudopotential inhibits formation of the Cooper pairs in superconductors by reducing the effective electron-phonon coupling constant [138-140]. We have shown the pressure dependent values of $N(E_F)$ and $\mu^*$ in Table 11, below.

**Table 11**

The total density of states at the Fermi level, $N(E_F)$ (states/eV.unit cell) and Coulomb pseudopotential, $\mu^*$ of CsV$_3$Sb$_5$ under different hydrostatic pressures (GPa).

| P | $N(E_F)$ | $\mu^*$ |
|---|---|---|
| 0 | 5.35 | 0.219 |
| 5 | 5.24 | 0.218 |
| 10 | 5.90 | 0.222 |
| 15 | 5.63 | 0.221 |
| 20 | 4.49 | 0.213 |

There is weak variation of the TDOS at the Fermi level and Coulomb pseudopotential with pressure as seen in Table 11.

### 3.6.3. Fermi surface topology

Understanding the Fermi surface (FS) is helpful to predict the electrical, magnetic, thermal, and optical properties of metals and semi-metals. Electrons near the Fermi sheets are involved in the superconducting state formation [141]. The Fermi surface of CsV$_3$Sb$_5$ Kagome compound under different pressures were constructed from the relevant band structures and are depicted in Figs. 9-13. The topology contains both electron- and hole-like sheets. Electron- and hole-like sheets have significant roles in determining the sign of the Hall coefficient of a material [108]. In the ground state, a cylinder-like FS is centered around the $\Gamma$-point (or along the $\Gamma$-A path) for band 37, which implies a strong two-dimensional nature of the electronic properties. A large and complex electron/hole-like hexagonal sheet is also found centered around the $\Gamma$-point. A tiny hole-like and electron-like sheet appears along the $\Gamma$-M path and around K-points, respectively, for band 38.

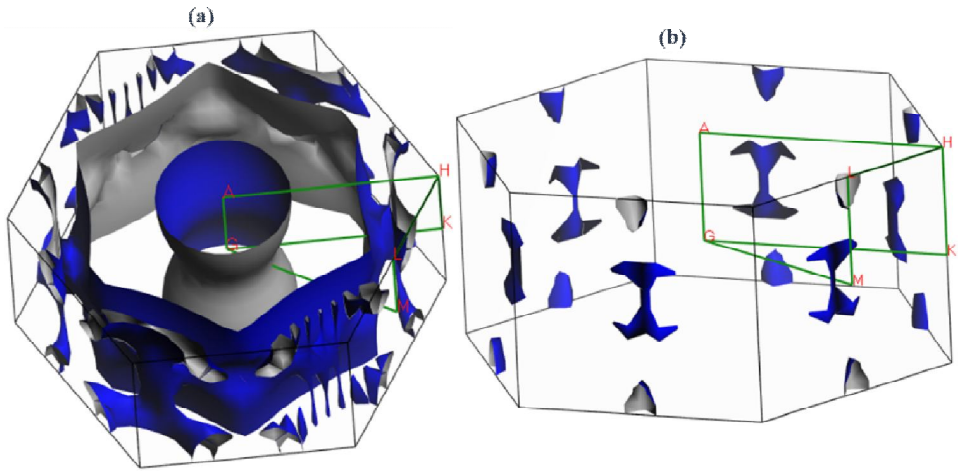

**Fig. 9.** The Fermi surfaces of CsV$_3$Sb$_5$ for band (a) 37 and (b) 38 at 0 GPa.

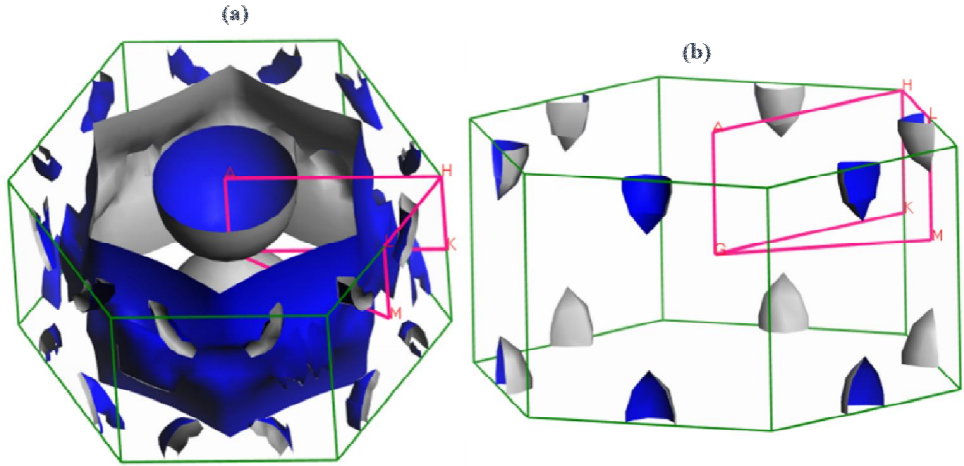

**Fig. 10.** The Fermi surfaces of $CsV_3Sb_5$ for band (a) 37 and (b) 38 at 5 GPa.

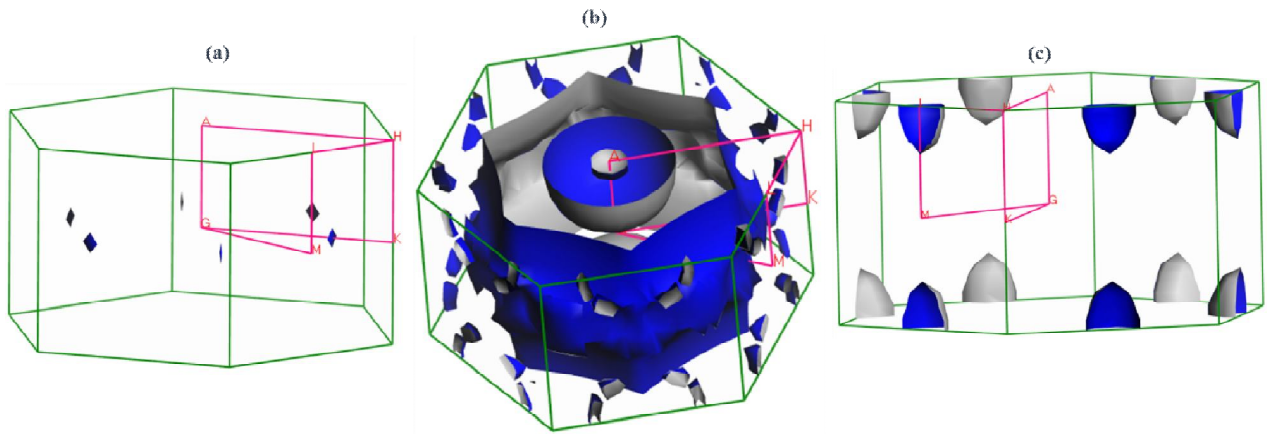

**Fig. 11.** The Fermi surfaces of $CsV_3Sb_5$ for band (a) 36, (b) 37, and (c) 38 at 10 GPa.

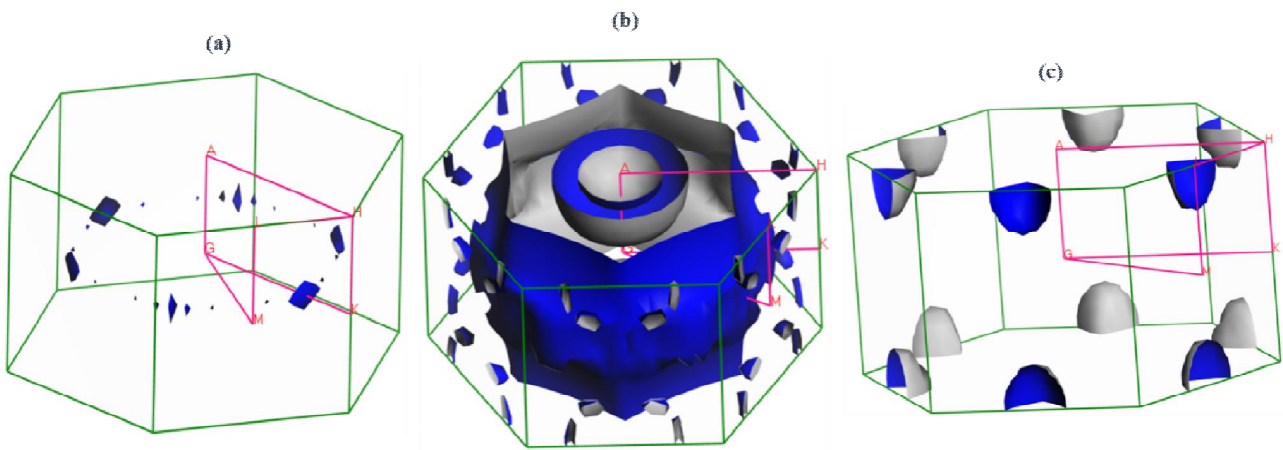

**Fig. 12.** The Fermi surfaces of $CsV_3Sb_5$ for band (a) 36, (b) 37, and (c) 38 at 15 GPa.

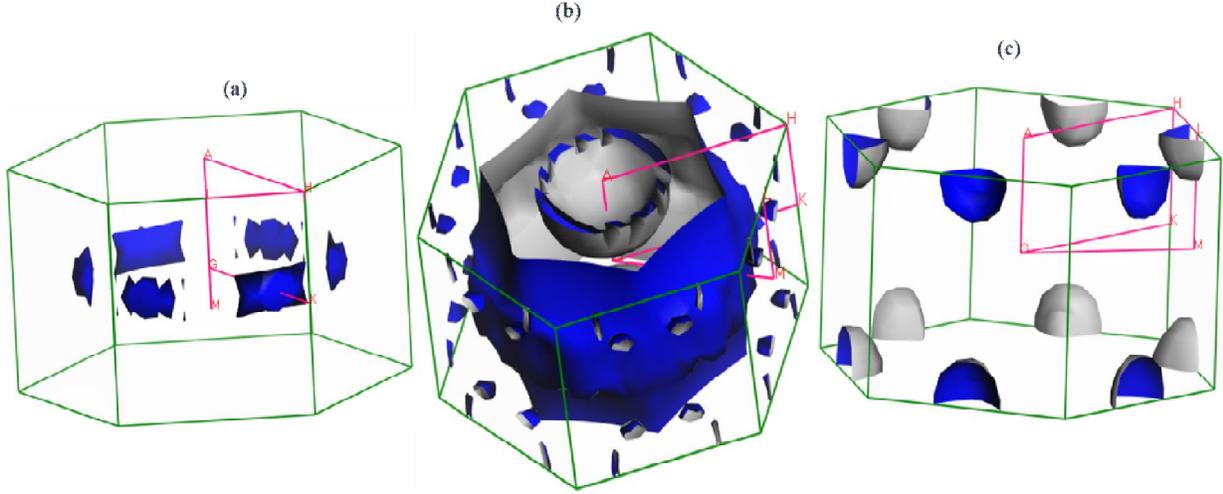

**Fig. 13.** The Fermi surfaces of $CsV_3Sb_5$ for band (a) 36, (b) 37, and (c) 38 at 20 GPa.

There is significant change in the Fermi surface topology with increasing pressure. As pressure increases, the contribution of band 38 increases in the overall Fermi surface. Band 38 creates electron pockets with hexagonal symmetry. For pressures above 5 GPa, another band (band 36) crosses the Fermi level and contributes to the Fermi surface. This leads to six disjointed Fermi sheets with complex structure (Figs. 11-13). The large Fermi sheet centered on the $\Gamma$-point has portions almost parallel to each other at all pressures leading to Fermi surface nesting. This feature often leads to formation of charge density waves (CDW) in strongly correlated electronic systems.

### 3.7. Optical properties

The study of optical properties of solids, one of the most fundamental factors, has been drawing great interest in different fields of current science and technology, such as in display devices, lasers, sensors, photo-electrodes, photonics, photo-detectors, solar cells, etc. It is also essential to consider optical anisotropy since some popular optical devices, like LCD screens, 3D movie screens, polarizers, and wave plates, are developed based on it [62]. Therefore, we studied the optical properties of $CsV_3Sb_5$ as a function of photon energies up to 20 eV for two different polarization directions [100] and [001] for the first time to get insight into the response to electromagnetic radiation. The Drude terms (damping of 0.05 eV) with an energy smearing of 0.5 eV have been used due to the metallic nature of the compound under study. All the frequency-dependent optical constants of the material, including the absorption coefficient $\alpha(\omega)$, reflectivity $R(\omega)$, optical conductivity $\sigma(\omega)$, refractive index $n(\omega)$ and $k(\omega)$, and energy loss function $L(\omega)$, can be explained from the complex dielectric function $\varepsilon(\omega)$. The intraband and interband transitions dominate the optical behavior of a metallic solid in the lower and higher energy regions, respectively [102]. In this section we have only shown the energy and electric field polarization dependent optical parameters of $CsV_3Sb_5$ in the ground state.

The frequency dependent real [$\varepsilon_1(\omega)$] and imaginary [$\varepsilon_2(\omega)$] parts of the dielectric function in the energy range 0-20 eV for $CsV_3Sb_5$ are depicted in Fig. 14(a). The real part, $\varepsilon_1(\omega)$, explains the electric polarization when light travels through a medium. On the contrary, the imaginary part, $\varepsilon_2(\omega)$, attributes to the electronic band structure and the absorption behavior of a material. The real part of the dielectric constant $\varepsilon_1$ at zero frequency describes the static dielectric constant $\varepsilon_1(0)$. Fig. 14(a) shows that the $\varepsilon_2(\omega)$ approaches zero at around 17.28 eV in the ultraviolet energy region. The absorption of photon by a medium takes place in the energy region where $\varepsilon_2(\omega)$ is nonzero. Both real and imaginary parts show anisotropy in the low energy region. The optical anisotropy in the dielectric function is moderate.

The index of refraction is an important parameter in determining the suitability of a material as optical devices, such as photonic crystals and waveguides. The complex refractive index of a material is expressed as: $N(\omega) = n(\omega) + ik(\omega)$, where $k(\omega)$ is the extinction coefficient. The extinction coefficient is a widely studied parameter for designing optoelectronic devices. The refractive index of $CsV_3Sb_5$ as a function of photon energy, obtained with two polarization directions, is depicted in Fig. 14(b). The low energy refractive index in the visible region is quite high. This makes the compound useful as wave guide and photonic engineering. The extinction coefficient shows peaks in the near ultraviolet (UV) region. Similar peaks are found in the imaginary part of the dielectric function at the same energy. This is expected since both these optical parameters are related to the loss of photon energy.

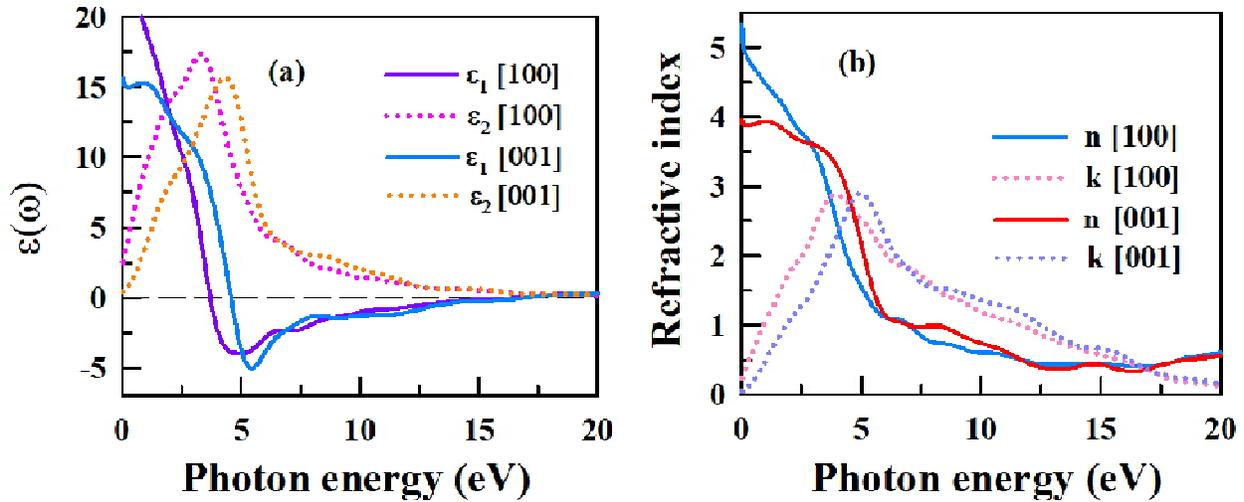

**Fig. 14.** Photon energy dependence of (a) real part, $\varepsilon_1(\omega)$ and imaginary part, $\varepsilon_2(\omega)$ of the dielectric functions and (b) refractive index of $CsV_3Sb_5$.

Fig. 15(a) illustrates the absorption spectrum ($\alpha$) of the sample as a function of photon energy. Like any other metallic system, the absorption starts from zero photon energy for both the polarization directions. This result is also consistent with the band structure and DOS studies (see section 3.6). The highest absorption of light appears around 7.30 eV and 5.37 eV for [100] and [001] polarization directions, respectively. A drastic reduction in absorption happens at an energy ~15 eV. Overall, light absorption is higher for the polarization direction [001]. The optical conductivity of a material is another widely studied parameter that signifies its electrical conductivity in the presence of an alternating electric field. The photoconductivity (σ) of $CsV_3Sb_5$ for both [100] and [001] polarizations is shown in Fig. 15(b). The photoconductivity, which starts from zero photon energy, manifests the absence of an electronic band gap. The conductivity increases and shifts toward higher energy for [001] direction of polarization. Optical conductivity becomes isotropic above 12 eV. The maximum photoconductivity of the compound is found at 3.7 eV and 4.5 eV for [100] and [001] polarizations, respectively.

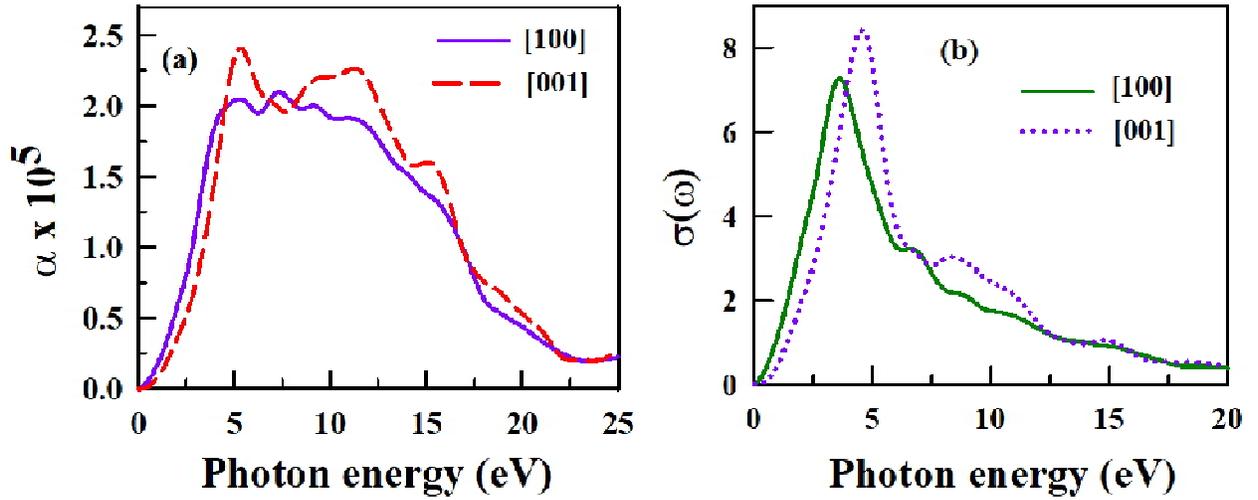

**Fig. 15.** Photon energy dependence of (a) absorption coefficient, $α(ω)$ and (b) photoconductivity, $σ(ω)$ of $CsV_3Sb_5$.

The reflectivity (*R*) of a material, an essential optical parameter, is defined as the ratio of the reflected to incident photon energy. It also provides information regarding the electronic structure of materials. The reflectivity spectrum of the compound is displayed in Fig. 16(a). The reflectivity starts from zero photon energy with 45% and 36% along [100] and [001] polarization directions, respectively (See Fig. 16(a)). The reflectivity spectrum shows clear optical anisotropy. R(ω) raises to its maximum values of 53% and 56% for [100] and [001] in the UV region, respectively. This implies that $CsV_3Sb_5$ can be used as an efficient reflector of solar radiation. The reflectivity spectrum falls sharply at incident photon energy of ~15 eV. Fig. 16(b) shows the electron energy loss spectrum as a function of frequency for $CsV_3Sb_5$. The bulk plasma frequency ($ω_P$) can be obtained from the energy loss spectrum since the peak in $L(ω)$ is associated with the plasma resonance and the corresponding frequency is called the plasma frequency ($ω_P$) [142]. The energy loss peak appears in the high energy region when $ε_2 < 1$ and $ε_1 = 0$ [34]. The peaks in $L(ω)$ appear at 17.37 eV for [100] and 17.05 eV for [001] polarization directions of the electric field. For incident frequencies higher than the plasma frequency, materials become transparent and exhibit the optical features of an insulator.

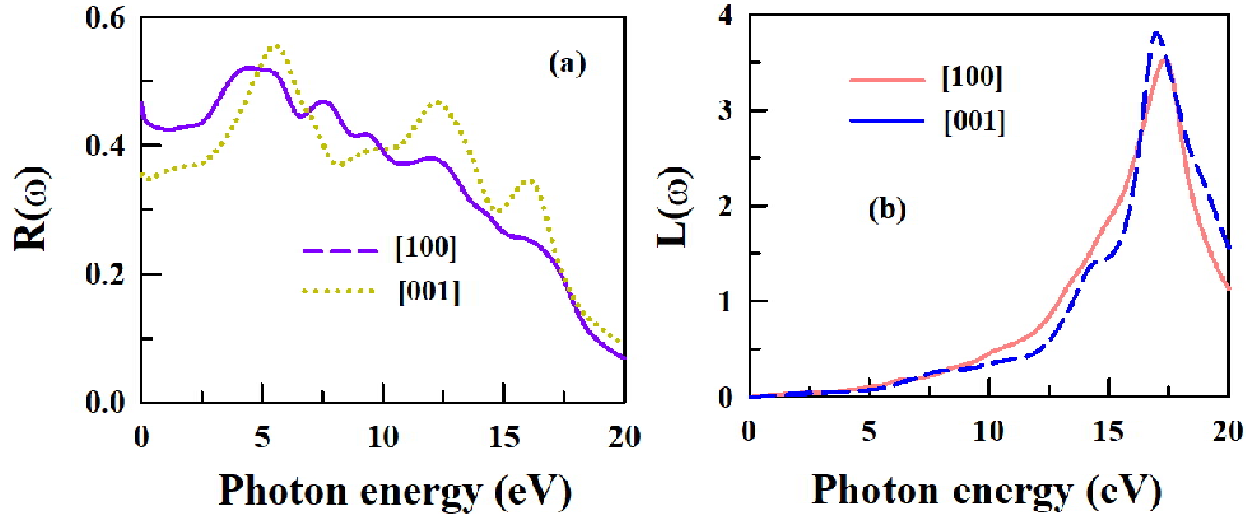

**Fig. 16.** Photon energy dependence of (a) reflection coefficient, $R(\omega)$ and (b) loss function, $L(\omega)$ of $CsV_3Sb_5$.

### 3.8. Bonding character
### 3.8.1 Electron density distribution

The nature of interatomic chemical bonds can be understood better by studying the electronic charge density around the atoms in a crystal. It displays the accumulation and depletion of charges around different atomic species of a compound. The covalent bonding between two atoms can be predicted from the charge accumulation between them, whereas the negative and positive charge balance at the atom positions indicates ionic bonding. Therefore, we have studied the electronic charge density distribution as depicted in Figs. 17(a-b). The adjacent color scale illustrates the intensity of total electronic charge density (e/Å$^3$). The blue and red colors represent the highest and the lowest charge densities, respectively. The charge density value ranges from 0.007 to 16.33 (electrons). The maximum charge accumulation is visible around V atoms in both Figs. 17(a) and 17(b). The balance of negative and positive charges around atoms exhibits ionic bonding. No sharing of electrons is observable between atoms in both charge density distribution maps. Therefore, ionic bondings between V-Cs and V-Sb atoms and covalent bonding between V-V atoms are expected. These findings are in accord with both the electronic density of states (See section 3.6.2) and the Mulliken bond population (See section 3.8.1) analysis. Therefore, the bonding in $CsV_3Sb_5$ is expected to be a mixture of both ionic and covalent. The V atoms on every planes (not shown here) exhibit maximum accumulation charges compared with Cs and Sb atoms. The charge distribution on different planes of the compound shows the direction and plane dependency. Fig. 18 confirms covalent bonding between V-V and V-Sb atoms since there are no atoms on the plane but the accumulation of charges are clearly observable (arrows).

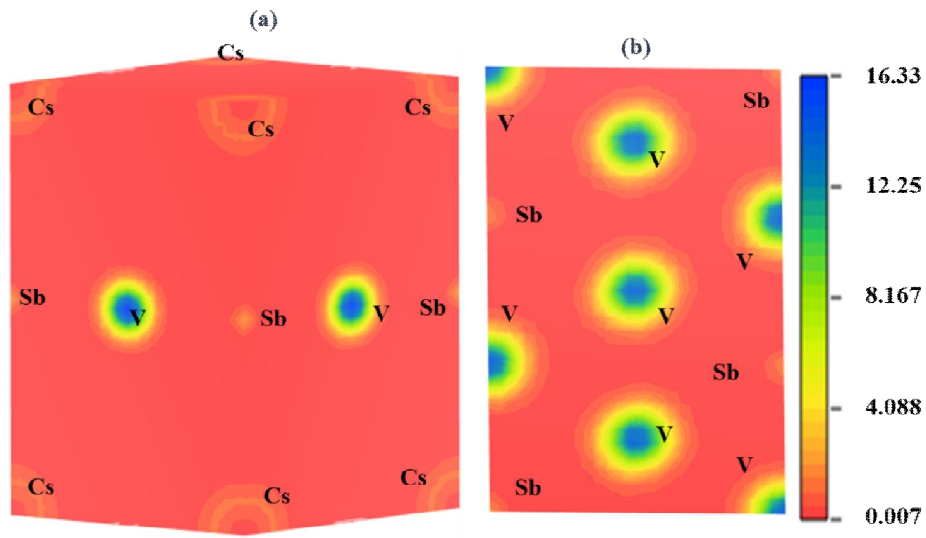

**Fig. 17.** Charge density distribution in various crystal planes. (a) 3D view and (b) in the (002) plane of CsV$_3$Sb$_5$ in the ground state. The charge density scale is displayed on the right.

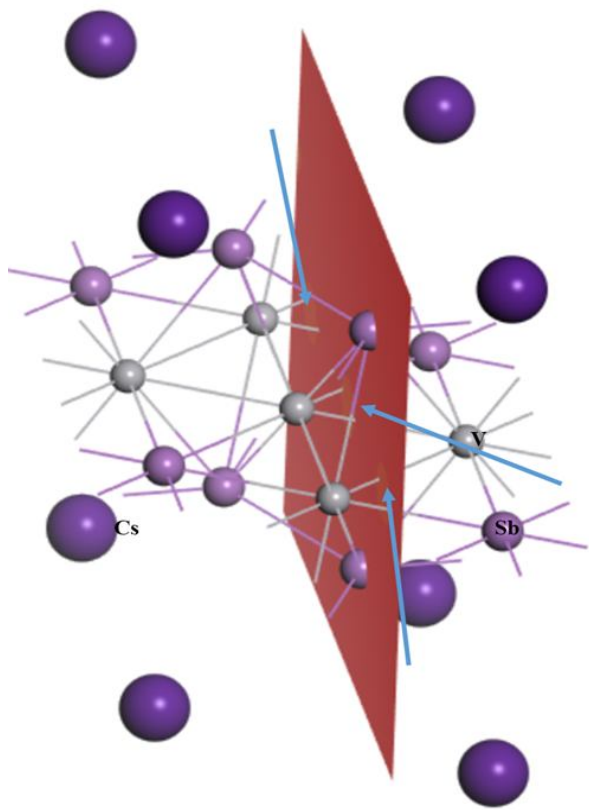

**Fig. 18.** Visualization of the presence of covalent bonds in CsV$_3$Sb$_5$ compound.

### 3.8.1. Bond population analysis

The bonding nature (ionic, covalent, and metallic), electric charge contribution from each atom, and hardness in $CsV_3Sb_5$ were further studied by applying the Mulliken bond population analysis [40] scheme. The charge spilling parameter and electric charges of Cs, V, and Sb species of the compound are summarized in Table 12. The calculation exhibits the total charge for V atoms is more than any other species the material contains. Sb1 possesses a slightly higher total charge than Sb2. The atomic charges of Cs, V, Sb1, and Sb2 atoms are 0.95e, -0.43e, 0.01e, and 0.29e, respectively. All the atoms deviated from their formal ionic charge expected for the purely ionic state (see Table 12). The positive and negative atomic charges refer to the electrons that are given away from and taken by the atoms, respectively. The calculated Mulliken charges predict that the charges transferred from V species to either Cs, Sb, or both. Therefore, an ionic contribution is presumed between V-Cs and V-Sb atoms. It also reflects the presence of covalent bonding between Cs-Cs, Cs-Sb, and Sb-Sb atoms since the number of electrons transferred by V atoms is not the same as the number of electrons received by any other species. Hence, the bonding behaviors in $CsV_3Sb_5$ are the combination of ionic and covalent bonds.

The difference between the formal ionic charge and Mulliken charge of an atom in a material is known as the effective valence charge (EVC) of that species [143]. It also exhibits the strength of either covalent, ionic, or both bonds. Zero EVC implies ideal ionic bonding, whereas all positive values measure the degree of covalency. The estimated EVCs for all species are given in Table 12. The nonzero EVC for each atom depicts prominent covalent bonding in $CsV_3Sb_5$ system. These results reinforce the electron density distribution and mechanical studies.

Despite being the most widely used method, the MPA sometimes provides results that conflict with the real chemical state due to its strong basis set dependency. As a result, the extended basis states can fail to provide a natural way to quantify the local atomic properties. On the contrary, the Hirshfeld population analysis (HPA) provides a more reliable data owing to its zero base set dependency [144]. As a result, we also studied the Hirshfeld charge of the material using the HPA, presented in Table 12. The atomic charges for each atoms, we got from HPA, are much smaller than those from the MPA. Although, both the HPA and the MPA agree on that electrons are transferred from Cs and Sb atoms to V atoms. The Cs atoms, most positively charged, in the system are strongly bound with the negatively charges V atoms. We also estimated EVC from the Hirshfeld charge and all the effective valences predict a higher level of covalency as compared to the EVC evaluated from the Mulliken charge.

**Table 12**
Charge spilling parameter (%), orbital charges (electron), atomic Mulliken charges (electron), formal ionic charge, EVC (electron), and Hirshfeld charge (electron) of $CsV_3Sb_5$ in the ground state.

| Compound | Charge spilling | Species | Mulliken atomic population | | | | Mulliken charge | Formal ionic charge | EVC | Hirshfeld charge | EVC |
|---|---|---|---|---|---|---|---|---|---|---|---|
| | | | s | P | d | Total | | | | | |
| $CsV_3Sb_5$ | 0.41 | Cs | 2.15 | 5.90 | 0.00 | 8.05 | 0.95 | +1 | 0.05 | 0.18 | 0.82 |
| | | V | 2.53 | 6.91 | 3.99 | 13.43 | -0.43 | +3 | 2.57 | -0.12 | 2.88 |
| | | Sb1 | 1.69 | 3.30 | 0.00 | 4.99 | 0.01 | 0 | 0.01 | 0.01 | 0.01 |
| | | Sb2 | 1.53 | 3.18 | 0.00 | 4.71 | 0.29 | 0 | 0.29 | 0.12 | 0.12 |

The bond type, bond overlap population, bond length, total number of each type of bond, pure covalent contribution, and total number of bonds in the bulk $CsV_3Sb_5$ are summarized in Table 13. The estimated Mulliken bond population represents the electron density constituting two atoms. Therefore, it can explain the strength of chemical bonds and the bond strength per unit volume. Table 13 shows the bond overlap populations for the nearest neighbors in the crystal. The zero (or close to zero) overlap population indicates the absence of any significant interaction between the electronic populations of the two atoms. Three V-Sb bonds in $CsV_3Sb_5$ shows zero overlap population (might be an indicator of ionic bonding between V-Sb atoms). The higher (lower) the positive bond population, a higher degree of covalency (iconicity). The positive bond overlap populations represent the states of bonding nature of interactions between the atoms involved [145]. On the other hand, the negative overlap populations indicate the existence of antibonding states or strong electrostatic repulsion between the atoms. The presence of bonding- and anti-bonding-type interactions is also confirmed in the compound. Out of 31 bonds in the unit cell, the antibonding states are constituted from the negative populations of V-Sb and Sb-Cs bonds. The antibonding interaction between atoms is the strongest between Sb-Cs, which is -2.62. The bond populations for Sb-Sb bonds, which only have a bonding nature, are much larger than any other bonds. The presence of large numbers of antibonding states could be an indicator of lower hardness. Therefore, there is a variety of chemical bonding in the crystal, such as metallic, covalent, and ionic bonds.

Phillips's homopolar band gap $E_h$ [146] characterizes the covalent bonding strength and is proportional to the activation energies of dislocation glide in polar covalent crystals. Covalency of the individual bonds can be determined by separating the average energy gap into covalent parts using the following equations [147]:

$$E_h = 39.74/d^{2.5} \qquad (45)$$

where, $d$ is the bond length/nearest neighbor distance (Å) and $E_h$ represents the pure covalent contribution (eV) to hardness. The estimated values are displayed in Table 13.

**Table 13**

The calculated Mulliken bond overlap population of $\mu$-type bond $P^\mu$, bond length $d^\mu$ (Å), pure covalent contribution $E_h$ (eV), total number of $\mu$-type bond $N^\mu$, bond volume $v_b^\mu$, and total number of bonds $\sum n^\mu$ for CsV$_3$Sb$_5$.

| Compound | Bond | | $P^\mu$ | $d^\mu$ | $E_h$ | $N^\mu$ | $v_b^\mu$ | $\Sigma n^\mu$ |
|---|---|---|---|---|---|---|---|---|
| CsV$_3$Sb$_5$ | V2-V3<br>V1-V3<br>V1-V2 | V-V | -0.26 | 2.727 | 3.236 | 3 | 4.018 | 31 |
| | V2-Sb5<br>V1-Sb5<br>V3-Sb5 | V-Sb | 0.00 | 2.727 | 3.236 | 3 | 4.018 | |
| | V3-Sb2<br>V3-Sb1<br>V2-Sb2<br>V2-Sb1<br>V3-Sb4<br>V3-Sb3<br>V2-Sb3<br>V2-Sb4<br>V1-Sb2<br>V1-Sb1<br>V1-Sb4<br>V1-Sb3 | V-Sb | -0.23 | 2.766 | 3.123 | 12 | 4.193 | |
| | Sb3-Sb4<br>Sb1-Sb2 | Sb-Sb | 0.17 | 3.149 | 2.258 | 2 | 6.187 | |
| | Sb2-Sb5<br>Sb1-Sb5<br>Sb4-Sb5<br>Sb3-Sb5 | Sb-Sb | 0.22 | 3.884 | 1.337 | 4 | 11.61 | |
| | Sb4-Cs1<br>Sb3-Cs1<br>Sb2-Cs1<br>Sb1-Cs1 | Sb-Cs | -2.62 | 3.924 | 1.303 | 4 | 11.98 | |
| | Sb2-Sb3<br>Sb1-Sb4 | Sb-Sb | 0.02 | 4.547 | 0.901 | 2 | 18.63 | |
| | Sb5-Cs1 | Sb-Cs | -2.38 | 4.615 | 0.869 | 1 | 19.47 | |

## 4. Conclusions

In this work, we have presented a comprehensive study of the synthesized kagome compound $CsV_3Sb_5$ following the DFT based first-principles calculations. The elastic, mechanical, chemical/bonding (charge density distribution, hardness, Mulliken bond population analysis, electronic (band, DOS, and Fermi surface topology), acoustic, thermophysical, optical properties are explored under different pressures for the very first time. At 0 GPa hydrostatic pressure, the calculated elastic constants of the compound ensure mechanical stability. The computed phonon dispersion confirms dynamical instability in the ground state. The compound possesses a moderate level of mechanical anisotropy, and hardness. All the mechanical, structural, and elastic parameters indicate onset of lattice instability for pressures in the 16 – 18 GPa range. The compound is highly machinable with an excellent dry lubricating prospect [57,58]. At the same time the Kagome compound is ductile in nature which makes it attractive for engineering applications. The mixed bonding nature of ionic and covalent was found in $CsV_3Sb_5$. The Dirac points are observed in the band structure close to the Fermi level. The electronic bands crossing the Fermi level and finite total density of states at $E_F$ confirms metallic character of the $CsV_3Sb_5$. Number of bands crossing the Fermi level increases with increasing pressure. The total DOS at $E_F$ decreases with pressures. The estimated Coulomb pseudopotential of $CsV_3Sb_5$ implies strong electronic correlations. The significant hybridization exists between the V 3$d$ and Sb 5$p$ electronic orbitals. The charge density distribution exhibits direction dependency. Both the electron density distribution and Mullikan atomic population studies confirm maximum charge is transferred from the V atoms. The comparatively lower value of the universal log-Euclidean index ($A^L$) for $CsV_3Sb_5$ forecasts non-layered characteristics, although several earlier literature indicated layered nature in this material [12,15,148-151]. The Fermi surface topology changes with pressure. Distinct hole pockets with hexagonal symmetry appears due to the application of the pressure in the BZ. The Fermi surface contains both electron- and hole-like sheets. There is also a tendency towards nesting in the Fermi surface structure. The optical properties of the compound are also studied and reveal some interesting properties. Like band structure and DOS features, the optical spectra also reveal metallic nature. The compound under study has moderate optical anisotropy and high absorption capability of the UV light. The compound can also be used as an efficient reflector of solar radiation. All the thermal properties show anomaly in the pressure range 16 – 18 GPa. The Debye temperature, melting temperature, and phonon thermal conductivity are low in the ground state and they increase with pressure up to 10 GPa.

In summary, we have investigated a large number of physical properties of the Kagome compound $CsV_3Sb_5$ in this paper. The elastic, band structure, Fermi surface topology, acoustic, thermophysical properties with pressure, bonding, and optical properties at 0 GPa of the compound were investigated in-depth for the first time. We are hopeful that these results will stimulate researchers to investigate the material in further detail, both theoretically and experimentally.

## Acknowledgements

Authors are grateful to the Department of Physics, Chittagong University of Engineering & Technology (CUET), Chattogram-4349, Bangladesh, for providing the computing facilities for this work. This work was also carried out with the aid of a grant (grant number: 21-378 RG/PHYS/AS_G-FR3240319526) from UNESCO-TWAS and the Swedish International Development Cooperation Agency (SIDA). The views expressed herein do not necessarily represent those of UNESCO-TWAS, SIDA, or its Board of Governors.

## Data availability

The data sets generated and/or analyzed in this study are available from the corresponding author on reasonable request.

## Author Contributions

**M. I. Naher:** Methodology, Formal analysis, Writing − original draft; **M. A. Ali:** Formal analysis, Writing − original draft, Validation; **M. M. Hossain:** Formal analysis, Writing review & editing, Validation; **M. M. Uddin:** Writing review & editing, Validation; **S. H. Naqib:** Conceptualization, Formal analysis, Writing review & editing, Validation, Supervision.

## Competing Interests

The authors declare that they have no known competing financial interests or personal relationships that could have appeared to influence the work reported in this paper.

## References

1. B.R. Ortiz, L.C. Gomes, J.R. Morey, M. Winiarski, M. Bordelon, J.S. Mangum, I.W.H. Oswald, J.A. Rodriguez-Rivera, J.R. Neilson, S.D. Wilson, E. Ertekin, T.M. McQueen, E.S. Toberer, New kagome prototype materials: discovery of $KV_3Sb_5$, $RbV_3Sb_5$, and $CsV_3Sb_5$. Phys. Rev. Materials 3 (2019) 094407. https://doi.org/10.1103/PhysRevMaterials.3.094407
2. F.H. Yu, T. Wu, Z.Y. Wang, B. Lei, W.Z. Zhuo, J.J. Ying, X.H. Chen, Concurrence of anomalous Hall effect and charge density wave in a superconducting topological kagome metal. Phys. Rev. B 104 (2021) L041103. https://doi.org/10.1103/PhysRevB.104.L041103
3. M. Kang, S. Fang, J.-K. Kim, B.R. Ortiz, S.H. Ryu, J. Kim, J. Yoo, G. Sangiovanni, D.D. Sante, B.-G. Park, C. Jozwiak, A. Bostwick, E. Rotenberg, E. Kaxiras, S.D. Wilson, J.-H. Park, R. Comin, Twofold van Hove singularity and origin of charge order in topological kagome superconductor $CsV_3Sb_5$. Nature Physics 18 (2022) 301. https://doi.org/10.1038/s41567-021-01451-5
4. B.R. Ortiz, S.M.L. Teicher, Y. Hu, J.L. Zuo, P.M. Sarte, E.C. Schueller, A.M. Milinda Abeykoon, M.J. Krogstad, S. Rosenkranz, R. Osborn, R. Seshadri, L. Balents, J. He, S.D. Wilson, $CsV_3Sb_5$: A $\square_2$ Topological Kagome Metal with a Superconducting Ground State. Phys. Rev. Lett. 125 (2020) 247002. https://doi.org/10.1103/PhysRevLett.125.247002

5. T. Neupert, M.M. Denner, J.-X. Yin, R. Thomale, M.Z. Hasan, Charge order and superconductivity in kagome materials. Nature Physics 18 (2022) 137. https://doi.org/10.1038/s41567-021-01404-y
6. E. Liu, Y. Sun, N. Kumar, L. Muechler, A. Sun, L. Jiao, S.-Y. Yang, D. Liu, A. Liang, Q. Xu, J. Kroder, V. Süß, H. Borrmann, C. Shekhar, Z. Wang, C. Xi, W. Wang, W. Schnelle, S. Wirth, Y. Chen, S.T.B. Goennenwein, C. Felser, Giant anomalous Hall effect in a ferromagnetic kagome-lattice semimetal. Nature Physics 14 (2018) 1125. https://doi.org/10.1038/s41567-018-0234-5
7. L. Ye, M. Kang, J. Liu, F. von Cube, C.R. Wicker, T. Suzuki, C. Jozwiak, A. Bostwick, E. Rotenberg, D.C. Bell, L. Fu, R. Comin, J.G. Checkelsky, Massive Dirac fermions in a ferromagnetic kagome metal. Nature 555 (2018) 638. https://doi.org/10.1038/nature25987
8. F. Du, S. Luo, R. Li, B.R. Ortiz, Y. Chen, S.D. Wilson, Y. Song, H. Yuan, Evolution of superconductivity and charge order in pressurized $RbV_3Sb_5$. Chinese Phys. B 31 (2022) 017404. DOI 10.1088/1674-1056/ac4232
9. Q. Yin, Z. Tu, C. Gong, Y. Fu, S. Yan, H. Lei, Superconductivity and normal-state properties of kagome metal $RbV_3Sb_5$ ingle crystals. Chin. Phys. Lett. 38 (2021) 037403. doi: 10.1088/0256-307X/38/3/037403
10. B.R. Ortiz, P.M. Sarte, E.M. Kenney, M.J. Graf, S.M.L. Teicher, R. Seshadri, S.D. Wilson, Superconductivity in the $Z_2$ kagome metal $KV_3Sb_5$. Phys. Rev. Materials 5 (2021) 034801. https://doi.org/10.1103/PhysRevMaterials.5.034801
11. F. Du, S. Luo, B.R. Ortiz, Y. Chen, W. Duan, D. Zhang, X. Lu, S.D. Wilson, Y. Song, H. Yuan, Pressure-induced double superconducting domes and charge instability in the kagome metal $KV_3Sb_5$. Phys. Rev. B 103 (2021) L220504. https://doi.org/10.1103/PhysRevB.103.L220504
12. F.H. Yu, D.H. Ma, W.Z. Zhuo, S.Q. Liu, X.K. Wen, B. Lei, J.J. Ying, X.H. Chen, Unusual competition of superconductivity and charge-density-wave state in a compressed topological kagome. Metal. Nat. Commun. 12 (2021) 3645. https://doi.org/10.1038/s41467-021-23928-w
13. C.C. Zhao, L.S. Wang, W. Xia, Q.W. Yin, J.M. Ni, Y.Y. Huang, C.P. Tu, Z.C. Tao, Z.J. Tu, C.S. Gong, H.C. Lei, Y.F. Guo, X.F. Yang, S.Y. Li, Nodal superconductivity and superconducting domes in the topological Kagome metal $CsV_3Sb_5$ (2021). Preprint at https://arxiv.org/abs/2102.08356.
14. N.N. Wang, K.Y. Chen, Q.W. Yin, Y.N.N. Ma, B.Y. Pan, X. Yang, X.Y. Ji, S.L. Wu, P.F. Shan, S.X. Xu, Z.J. Tu, C.S. Gong, G.T. Liu, G. Li, Y. Uwatoko, X.L. Dong, H.C. Lei, J.P. Sun, J.-G. Cheng, Competition between charge-density-wave and superconductivity in the kagome metal $RbV_3Sb_5$. Physical Review Research 3 (2021) 043018. https://doi.org/10.1103/PhysRevResearch.3.043018
15. S. Cho, H. Ma, W. Xia, Y. Yang, Z. Liu, Z. Huang, Z. Jiang, X. Lu, J. Liu, Z. Liu, J. Li, J. Wang, Y. Liu, J. Jia, Y. Guo, J. Liu, D. Shen, Emergence of New van Hove Singularities in the Charge Density Wave State of a Topological Kagome Metal RbV$_3$Sb$_5$. Phys. Rev. Lett. 127 (2021) 236401. https://doi.org/10.1103/PhysRevLett.127.236401

16. Z. Zhang, Z. Chen, Y. Zhou, Y. Yuan, S. Wang, J. Wang, H. Yang, C. An, L. Zhang, X. Zhu, Y. Zhou, X. Chen, J. Zhou, Z. Yang, Pressure-induced reemergence of superconductivity in the topological kagome metal CsV$_3$Sb$_5$. Phys. Rev. B 103 (2021) 224513. https://doi.org/10.1103/PhysRevB.103.224513

17. J.-F. Zhang, K. Liu, Z.-Y. Lu, First-principles study of the double-dome superconductivity in the kagome material CsV$_3$Sb$_5$ under pressure. Phys. Rev. B 104 (2021) 195130. https://doi.org/10.1103/PhysRevB.104.195130

18. X. Chen, X. Zhan, X. Wang, J. Deng, X.-B. Liu, X. Chen, J.-G. Guo, X. Chen, Highly robust reentrant superconductivity in CsV$_3$Sb$_5$ under pressure. Chin. Phys. Lett. 38 (2021) 057402. doi:10.1088/0256-307X/38/5/057402

19. E. Uykur, B.R. Ortiz, O. Iakutkina, M. Wenzel, S.D. Wilson, M. Dressel, A.A. Tsirlin, Low-energy optical properties of the nonmagnetic kagome metal CsV$_3$Sb$_5$. Phys. Rev. B 104 (2021) 045130. https://doi.org/10.1103/PhysRevB.104.045130

20. X. Zhou, Y. Li, X. Fan, J. Hao, Y. Dai, Z. Wang, Y. Yao, H.-H. Wen, Origin of charge density wave in the kagome metal CsV$_3$Sb$_5$ as revealed by optical spectroscopy. Phys. Rev. B104 (2021) L041101. https://doi.org/10.1103/PhysRevB.104.L041101

21. Y. Fu, N. Zhao, Z. Chen, Q. Yin, Z. Tu, C. Gong, C. Xi, X. Zhu, Y. Sun, K. Liu, H. Lei, Quantum Transport Evidence of Topological Band Structures of Kagome Superconductor CsV$_3$Sb$_5$. Phys. Rev. Lett. 127 (2021) 207002. https://doi.org/10.1103/PhysRevLett.127.207002

22. H. Tan, Y. Liu, Z. Wang, B. Yan, Charge Density Waves and Electronic Properties of Superconducting Kagome Metals. Phys. Rev. Lett. 127 (2021) 046401. https://doi.org/10.1103/PhysRevLett.127.046401

23. S.J. Clark, M.D. Segall, C.J. Pickard, P.J. Hasnip, M.J. Probert, K. Refson, M.C. Payne, First principles methods using CASTEP. Z. Kristallogr 220 (2005) 567. https://doi.org/10.1524/zkri.220.5.567.65075

24. R.G. Parr, Density Functional Theory. Ann. Rev. Phys. Chern. 34 (1983) 631. https://doi.org/10.1146/annurev.pc.34.100183.003215

25. Materials studio CASTEP manual © Accelrys2010. http://www.tcm.phy.cam.ac.uk/castep/documentation/WebHelp/CASTEP.html

26. J.P. Perdew, K. Burke, M. Ernzerhof, Generalized Gradient Approximation Made Simple. Phys. Rev. Lett. 77 (1996) 3865. https://doi.org/10.1103/PhysRevLett.77.3865

27. J.P. Perdew, A. Ruzsinszky, G.I. Csonka, O.A. Vydrov, G.E. Scuseria, L.A. Constantin, X. Zhou, K. Burke, Restoring the Density-Gradient Expansion for Exchange in Solids and Surfaces. Phys. Rev. Lett. 100 (2008) 136406. https://doi.org/10.1103/PhysRevLett.100.136406

28. M.D.L. Pierre, R. Orlando, L. Maschio, K. Doll, P. Ugliengo, R. Dovesi, Performance of six functionals (LDA, PBE, PBESOL, B3LYP, PBE0, and WC1LYP) in the simulation of vibrational and dielectric properties of crystalline compounds. The case of forsterite $Mg_2SiO_4$. Journal of Computational Chemistry 32 (2011) 1775. https://doi.org/10.1002/jcc.21750

29. D. Vanderbilt, Soft self-consistent pseudopotentials in a generalized eigenvalue formalism. Phys. Rev. B 41 (1990) 7892. https://doi.org/10.1103/PhysRevB.41.7892
30. T.H. Fischer, J. Almlof, General methods for geometry and wave function optimization. J. Phys. Chem. 96 (1992) 9768. https://doi.org/10.1021/j100203a036
31. H. Monkhorst, J. Pack, Special points for Brillouin-zone integrations. Phys. Rev. B 13 (1976) 5188. https://doi.org/10.1103/PhysRevB.13.5188
32. O.H. Nielsen, R.M. Martin, First-principles calculation of stress. Phys. Rev. Lett. 50 (1983) 697. https://doi.org/10.1103/PhysRevLett.50.697
33. J.P. Watt, Hashin–Shtrikman bounds on the effective elastic moduli of polycrystals with orthorhombic symmetry. J. Appl. Phys. 50 (1979) 6290. https://doi.org/10.1063/1.325768
34. J.P. Watt, L. Peselnick, Clarification of the Hashin–Shtrikman bounds on the effective elastic moduli of polycrystals with hexagonal, trigonal, and tetragonal symmetries. J. Appl. Phys. 51 (1980) 1525. https://doi.org/10.1063/1.327804
35. J. Sun, H.T. Wang, J.L. He, Y.J. Tian, Ab initio investigations of optical properties of the high-pressure phases of ZnO. Phys. Rev. B 71 (2005) 125132. https://doi.org/10.1103/PhysRevB.71.125132
36. S. Saha, T.P. Sinha, Electronic structure, chemical bonding, and optical properties of paraelectric $BaTiO_3$. Phys. Rev. B 62 (2000) 8828. https://doi.org/10.1103/PhysRevB.62.8828
37. M.Q. Cai, Z. Yin, M.S. Zhang, First-principles study of optical properties of barium titanate. Appl. Phys. Lett. 83 (2003) 2805. https://doi.org/10.1063/1.1616631
38. D. Sanchez-Portal, E. Artacho, J.M. Soler, Projection of plane-wave calculations into atomic orbitals. Solid State Commun. 95 (1995) 685. https://doi.org/10.1016/0038-1098(95)00341-X
39. M.D. Segall, R. Shah, C.J. Pickard, M.C. Payne, Population analysis of plane-wave electronic structure calculations of bulk materials. Phys. Rev. B 54 (1996) 16317. https://doi.org/10.1103/PhysRevB.54.16317
40. R.S. Mulliken, Electronic Population Analysis on LCAO–MO Molecular Wave Functions. II. Overlap Populations, Bond Orders, and Covalent Bond Energies. J. Chem. Phys. 23 (1955) 1833. https://doi.org/10.1063/1.1740589
41. J. Zhao, W. Wu, Y. Wang, S. A. Yang, Electronic correlations in the normal state of kagome superconductor $KV_3Sb_5$. Phys. Rev. B 103 (2021) L241117. https://doi.org/10.1103/PhysRevB.103.L241117
42. A.J. Majewski, P. Vogl, Simple model for structural properties and crystal stability of sp-bonded solids. Phys. Rev. B 35 (1987) 9666. https://doi.org/10.1103/PhysRevB.35.9666

43. H. Zhao, H. Li, B.R. Ortiz, S.M.L. Teicher, T. Park, M. Ye, Z. Wang, L. Balents, S.D. Wilson, I. Zeljkovic, Cascade of correlated electron states in the kagome superconductor $CsV_3Sb_5$. Nature 599 (2021) 216. https://doi.org/10.1038/s41586-021-03946-w
44. N. Ratcliff, L. Hallett, B.R. Ortiz, S.D. Wilson, J.W. Harter, Coherent phonon spectroscopy and interlayer modulation of charge density wave order in the kagome metal $CsV_3Sb_5$. Phys. Rev. Materials 5 (2021) L111801. https://doi.org/10.1103/PhysRevMaterials.5.L111801
45. A. Subedi, Hexagonal-to-base-centered-orthorhombic 4Q charge density wave order in kagome metals $KV_3Sb_5$, $RbV_3Sb_5$, and $CsV_3Sb_5$. Phys. Rev. Materials 6 (2022) 015001. https://doi.org/10.1103/PhysRevMaterials.6.015001
46. M.I. Naher, S.H. Naqib, First-principles insights into the mechanical, optoelectronic, thermophysical, and lattice dynamical properties of binary topological semimetal $BaGa_2$. Results in Physics 37 (2022) 105507. https://doi.org/10.1016/j.rinp.2022.105507
47. M.I. Naher, M.A. Afzal, S.H. Naqib, A comprehensive DFT based insights into the physical properties of tetragonal superconducting $Mo_5PB_2$. Results in Physics 28 (2021) 10461. https://doi.org/10.1016/j.rinp.2021.104612
48. M.I. Naher, S.H. Naqib, A comprehensive study of the thermophysical and optoelectronic properties of $Nb_2P_5$ via ab-initio technique. Results in Physics 28 (2021) 104623. https://doi.org/10.1016/j.rinp.2021.104623
49. T. Lay, T.C. Wallace, Modern global seismology (Elsevier, 1995).
50. C.H. Turner, S.C. Cowin, J.Y. Rho, R.B. Ashman, J.C. Rice, The fabric dependence of the orthotropic elastic constants of cancellous bone. Journal of biomechanics 23 (1990) 549. https://doi.org/10.1016/0021-9290(90)90048-8
51. M. Born, K. Hang, Dynamical Theory and Experiments I, Springer-Verlag Publishers, Berlin, 1982.
52. M.I. Naher, S.H. Naqib, An ab-initio study on structural, elastic, electronic, bonding, thermal, and optical properties of topological Weyl semimetal Ta$X$ ($X$ = P, As). Sci. Rep. 11 (2021) 5592. https://doi.org/10.1038/s41598-021-85074-z
53. L. Vitos, P.A. Korzhavyi, B. Johansson, Stainless steel optimization from quantum mechanical calculations. Nat. Mater. 2 (2003) 25. https://doi.org/10.1038/nmat790
54. K.J. Puttlitz, K.A. Stalter, Handbook of Lead-free Solder Technology for Microelectronic Assemblies, vol. 98, CRC Press, New York, 2004. Springer.
55. M.J. Phasha, P.E. Ngoepe, H.R. Chauke, *et al.*, Link between structural and mechanical stability of fcc- and bcc-based ordered MgeLi alloys. Intermetallics 18 (2010) 2083. https://doi.org/10.1016/j.intermet.2010.06.015
56. Z. Sun, D. Music, R. Ahuja, J.M. Schneider, Theoretical investigation of the bonding and elastic properties of nanolayered ternary nitrides. Phys. Rev. B 71 (2005) 193402. https://doi.org/10.1103/PhysRevB.71.193402
57. C. Kittel, Introduction to Solid State Physics. 1996 7th edn. (New York: Wiley).

58. T. Reeswinkel, D. Music, J.M. Schneide, *Ab initio* calculations of the structure and mechanical properties of vanadium oxides. J. Phys.: Condens. Matter 21 (2009) 145404. DOI 10.1088/0953-8984/21/14/145404
59. W. Voigt, Lehrbuch der Kristallphysik. Leipzig: Taubner; 1928. p. 962.
60. A. Reuss, Berechnung der Fließgrenze von Mischkristallen auf Grund der Plastizitätsbedingung für Einkristalle. Z. Angew. Math. Mech. 9 (1929) 49. https://doi.org/10.1002/zamm.19290090104
61. R. Hill, The Elastic Behaviour of a Crystalline Aggregate. Proc. Phys. Soc. A 65 (1952) 349. DOI: 10.1088/0370-1298/65/5/307
62. M.I. Naher, S.H. Naqib, Possible applications of $Mo_2C$ in the orthorhombic and hexagonal phases explored via ab-initio investigations of elastic, bonding, optoelectronic and thermophysical properties. Results in Physics 37 (2022) 105505. https://doi.org/10.1016/j.rinp.2022.105505
63. K.A. Gschneidner, Physical properties and interrelationships of metallic and semimetallic elements. Solid State Phys. 16 (1964) 275. https://doi.org/10.1016/S0081-1947(08)60518-4
64. S.F. Pugh, Relation between the elastic moduli and the plastic properties of polycrystalline pure metals. Phil. Mag. 45 (1954) 823. https://doi.org/10.1080/14786440808520496
65. P. Ravindran, L. Fast, P. Korzhavyi, B. Johansson, Density functional theory for calculation of elastic properties of orthorhombic crystals: Application to $TiSi_2$. J. App. Phys. 84 (1998) 4891. https://doi.org/10.1063/1.368733
66. K. Tanaka, K. Okamoto, H. Inui, Y. Minonishi, M. Yamaguchi, M. Koiwa, Elastic constants and their temperature dependence for the intermetallic compound $Ti_3Al$. Philos. Mag. A 73 (1996) 1475. https://doi.org/10.1080/01418619608245145
67. G.N. Greaves, A.L. Greer, R.S. Lakes, T. Rouxel, Poisson's ratio and modern materials. Nat. Mater. 10 (2011) 823. https://doi.org/10.1038/nmat3134
68. W. Koster, H. Franz, Poisson's ratio for metals and alloys. Metall. Rev. 6 (1961) 1. https://doi.org/10.1179/mtlr.1961.6.1.1
69. H. Fu, D. Li, F. Peng, T. Gao, X. Cheng, Ab initio calculations of elastic constants and thermodynamic properties of NiAl under high pressures. Comput. Mater. Sci. 44 (2008) 774. https://doi.org/10.1016/j.commatsci.2008.05.026
70. P.H. Mott, J.R. Dorgan, C.M. Roland, The bulk modulus and Poisson's ratio of "incompressible" materials. Journal of Sound and Vibration 312 (2008) 572. https://doi.org/10.1016/j.jsv.2008.01.026
71. J. Haines, J.M. Leger, G. Bocquillon, Synthesis and design of superhard materials. Annu. Rev. Mater. Res. 31 (2001) 1. https://doi.org/10.1146/annurev.matsci.31.1.1
72. V.V. Bannikov, I.R. Shein, A.L. Ivanovskii, Electronic structure, chemical bonding and elastic properties of the first thorium-containing nitride perovskite $TaThN_3$, Phys. Stat. Sol. (RRL) 1 (2007) 89. https://doi.org/10.1002/pssr.200600116
73. D.G. Pettifor, Theoretical predictions of structure and related properties of intermetallics. Mater. Sci. Technol. 8 (1992) 345. https://doi.org/10.1179/mst.1992.8.4.345

74. X. Zeng, R. Peng, Y. Yu, Z. Hu, Y. Wen, L. Song, Pressure Effect on Elastic Constants and Related Properties of Ti$_3$Al Intermetallic Compound: A First-Principles Study. Materials 11 (2018) 2015. https://doi.org/10.3390/ma11102015.
75. D. Qu, C. Li, L. Bao, Z. Kong, Y. Duan, Structural, electronic, and elastic properties of orthorhombic, hexagonal, and cubic Cu$_3$Sn intermetallic compounds in Sn–Cu lead-free solder. J Phys. Chem. Solids 138 (2020) 109253. https://doi.org/10.1016/j.jpcs.2019.109253
76. W. Feng, S. Cui, Mechanical and electronic properties of Ti$_2$AlN and Ti$_4$AlN$_3$: a first-principles study. Can. J. Phys. 92 (2014) 1652. https://doi.org/10.1139/cjp-2013-0746
77. E. Kaxiras, M.S. Duesbery, Free energies of generalized stacking faults in Si and implications for the brittle-ductile transition. Phys. Rev. Lett. 70 (1993) 3752. https://doi.org/10.1103/PhysRevLett.70.3752
78. G. Lu, The Peierls-Nabarro model of dislocations: a venerable theory and its current development. In: Yip, S. (eds) Handbook of Materials Modeling. Springer, Dordrecht. 1 (2005). https://doi.org/10.1007/978-1-4020-3286-8_41
79. H. Siethoff, Homopolar band gap and thermal activation parameters of plasticity of diamond and zinc-blende semiconductors. Journal of Applied Physics 87 (2000) 3301. https://doi.org/10.1063/1.372340
80. J. Meneve, K. Vercammen, E. Dekempeneer, J. Smeets, Thin tribological coatings: magic or design? Surface and Coatings Technology 94-95 (1997) 476. https://doi.org/10.1016/S0257-8972(97)00430-1
81. J.H. Westbrook, H. Conrad (Eds.), The Science of Hardness testing and Its Research Applications, American Society for Metals, Metals Park, 1973.
82. S.P. Baker, R.F. Cook, S.G. Corcoran, N.R. Moody (Eds.), Fundamentals of Nanoindentation and Nanotribology II, Mater. Res. Soc. Symp. Proc. 649 (2001).
83. A. Kumar, W.J. Meng, Y.T. Cheng, J.S. Zabinski, G.L. Doll, S. Veprek (Eds.), Surface Engineering 2002-Synthesis, Characterization and Applications, Mater. Res. Soc. Symp. Proc. 750 (2003).
84. Y.T. Cheng, T. Page, G.M. Pharr, M. Swain, K.J. Wahl (Eds.), Fundamentals and Applications of Instrumented Indentation in Multidisciplinary Research, J. Mater. Res. 19 (2004) 1.
85. Z. Liu, M.G. Scanlon, Modeling Indentation of Bread Crumb by Finite Element Analysis. Biosyst. Eng. 85 (2003) 477. https://doi.org/10.1016/S1537-5110(03)00093-X
86. Y.-T. Cheng, C.-M. Cheng, Scaling, dimensional analysis, and indentation measurements. Mater. Sci. Eng. R 44 (2004) 91. https://doi.org/10.1016/j.mser.2004.05.001
87. F. Gao, J. He, E. Wu, S. Liu, D. Yu, D. Li, S. Zhang, Y. Tian, Hardness of Covalent Crystals. Phys. Rev. Lett. 91 (2003) 015502. https://doi.org/10.1103/PhysRevLett.91.015502

88. A. Simunek, J. Vackar, Hardness of Covalent and Ionic Crystals: First-Principle Calculations. Phys. Rev. Lett. 96 (2006) 085501. https://doi.org/10.1103/PhysRevLett.96.085501
89. V.A. Mukhanov, O.O. Kurakevych, V.L. Solozhenko, Thermodynamic aspects of materials' hardness: prediction of novel superhard high-pressure phases. High Press. Res. 28 (2008) 531. https://doi.org/10.1080/08957950802429052
90. K. Li, X. Wang, F. Zhang, D. Xue, Electronegativity Identification of Novel Superhard Materials. Phys. Rev. Lett. 100 (2008) 235504. https://doi.org/10.1080/08957950802429052
91. M.M. Smedskjaer, J.C. Mauro, Y. Yue, Prediction of Glass Hardness Using Temperature-Dependent Constraint Theory. Phys. Rev. Lett. 105 (2010) 115503. https://doi.org/10.1103/PhysRevLett.105.115503
92. A.L. Ivanovskii, Hardness of hexagonal $AlB_2$-like diborides of *s*, *p*, and *d* metals from semi-empirical estimations. International Journal of Refractory Metals and Hard Materials 36 (2013) 179. https://doi.org/10.1016/j.ijrmhm.2012.08.013
93. X.-Q. Chen, H. Niu, D. Li, Y. Li, Modeling hardness of polycrystalline materials and bulk metallic glasses. Intermetallics 19 (2011) 1275. https://doi.org/10.1016/j.intermet.2011.03.026
94. X. Jiang, J. Zhao, X. Jiang, Correlation between hardness and elastic moduli of the covalent crystals. Comput. Mater. Sci. 50 (2011) 2287. https://doi.org/10.1016/j.commatsci.2011.01.043
95. A. Erdemir, Modern Tribology Handbook, Vol. II, ed. B. Bhushan (CRC Press, Boca Raton, FL, 2001) p. 787.
96. K. Holmberg and A. Matthews, in: Coatings Tribology, ed. D. Dowson, (Elsevier, Netherlands, 1994) p.1.
97. C. Donnet, A. Erdemir, Solid Lubricant Coatings: Recent Developments and Future Trends. Tribology Letters 17 (2004) 389. https://doi.org/10.1023/B:TRIL.0000044487.32514.1d
98. A. Šimůnek, Anisotropy of hardness from first principles: The cases of $ReB_2$ and $OsB_2$. Phys. Rev. B 80 (2009) 060103(R). https://doi.org/10.1103/PhysRevB.80.060103
99. M.D. Jong, W. Chen, T. Angsten, *et al.*, Charting the complete elastic properties of inorganic crystalline compounds. Sci Data 2 (2015) 150009. https://doi.org/10.1038/sdata.2015.9
100. S.I. Ranganathan, M. Ostoja-Starzewski, Universal elastic anisotropy index. Phys. Rev. Lett. 101 (2008) 055504. https://doi.org/10.1103/PhysRevLett.101.055504
101. D.H. Chung, W.R. Buessem, in Anisotropy in Single Crystal Refractory Compound, Edited by F.W. Vahldiek and S.A. Mersol, Vol. 2 (Plenum press, New York, 1968).
102. M.I. Naher, S.H. Naqib, Structural, elastic, electronic, bonding, and optical properties of topological $CaSn_3$ semimetal. J. Alloys Compd. 829 (2020) 154509. https://doi.org/10.1016/j.jallcom.2020.154509

103. C.M. Kube, M.D. Jong, Elastic constants of polycrystals with generally anisotropic crystals. Journal of Applied Physics 120 (2016) 165105. https://doi.org/10.1063/1.4965867
104. V. Milman, M.C. Warren, Elasticity of hexagonal BeO. J. Phys.: Condens. Matter 13 (2001) 5585. DOI:10.1088/0953-8984/13/2/302
105. R. Gaillac, P. Pullumbi, F.X. Coudert, ELATE: an open-source online application for analysis and visualization of elastic tensors. Journal of Physics: Condensed Matter 28 (2016) 275201. DOI:10.1088/0953-8984/28/27/275201
106. NDT Education Resource Center. Introduction to Sound, 2011. Consulted in https://www.nde-ed.org/EducationRe sources/HighSchool/Sound/introsound.htm.
107. O.L. Anderson, A simplified method for calculating the Debye temperature from elastic constants. J. Phys. Chem. Solids 24 (1963) 909. https://doi.org/10.1016/0022-3697(63)90067-2
108. M.I. Naher, F. Parvin, A.K.M.A. Islam, S.H. Naqib, Physical properties of niobium-based intermetallics ($Nb_3B$; $B$ = Os, Pt, Au): a DFT-based ab-initio study. Eur. Phys. J. B 91 (2018) 289. https://doi.org/10.1140/epjb/e2018-90388-9
109. E.L. Caudana, O. Quiroz, A. Rodrìguez, L. Yèpez, D. Ibarra, Classification of materials by acoustic signal processing in real time for NAO robots. International Journal of Advanced Robotic Systems (2017). https://doi.org/10.1177/1729881417714996
110. E. Yasar, Y. Erdogan, Correlating sound velocity with the density, compressive strength and Young's modulus of carbonate rocks. International Journal of Rock Mechanics & Mining Sciences 41 (2004) 871. https://doi.org/10.1016/j.ijrmms.2004.01.012
111. S. Hughes, Medical ultrasound imaging. Physics Education 36 (2001) 468. DOI:10.1088/0031-9120/36/6/304
112. M.F. Ashby, Materials Selection in Mechanical Design (fourth edition) (2011). ISBN 978-1-85617-663-7
113. M.F. Ashby, P.J. Ferreira, D.L. Schodek (2009). Material Classes, Structure, and Properties. Nanomaterials, Nanotechnologies and Design, p-143. DOI:10.1016/b978-0-7506-8149-0.00006-4
114. E. Fouilhe, A. Houssay, I. Brémaud. Dense and hard woods in musical instrument making: Comparison of mechanical properties and perceptual "quality" grading. Acoustics 2012, Apr 2012, Nantes, France. pp.1-6. ffhal-00808368ff
115. B.D. Sanditov, S.B. Tsydypov, D.S. Sanditov, Relation between the Grüneisen Constant and Poisson's Ratio of Vitreous Systems. Acoustical Physics 53 (2007) 594. https://doi.org/10.1134/S1063771007050090
116. G.A. Slack, The thermal conductivity of nonmetallic crystals. Solid state Physics 34 (1979) 1. https://doi.org/10.1016/S0081-1947(08)60359-8
117. T.H.K. Barron, Grüneisen parameters for the equation of state of solids. Annals of Physics 1 (1957) 77. https://doi.org/10.1016/0003-4916(57)90006-4

118. Y. Yun, D. Legut, P.M. Oppeneer, Phonon spectrum, thermal expansion and heat capacity of $UO_2$ from first-principles. J. Nucl. Mater. 426 (2012) 109. https://doi.org/10.1016/j.jnucmat.2012.03.017
119. A.N. Kolmogorov, M. Calandra, S. Curtarolo, Thermodynamic stabilities of ternary metal borides: An ab initio guide for synthesizing layered superconductors. Phys. Rev. B 78 (2008) 094520. https://doi.org/10.1103/PhysRevB.78.094520
120. A.S. Disa, T.F. Nova, A. Cavalleri, Engineering crystal structures with light. Nature Physics 17 (2021) 1087. https://doi.org/10.1038/s41567-021-01366-1
121. G. Kresse, J. Furthmüller, J. Hafner, Ab initio Force Constant Approach to Phonon Dispersion Relations of Diamond and Graphite. Europhys. Lett. 32 (1995) 729. DOI: 10.1209/0295-5075/32/9/005
122. K. Parlinski, Z.Q. Li, Y. Kawazoe, First-Principles Determination of the Soft Mode in Cubic $ZeO_2$. Phys. Rev. Lett. 78 (1997) 4063. https://doi.org/10.1103/PhysRevLett.78.4063
123. A. Amon, E. Svanidze, R. Cardoso-Gil, M.N. Wilson, *et al.*, Noncentrosymmetric superconductor BeAu. Phys. Rev. B 97 (2018) 014501. https://doi.org/10.1103/PhysRevB.97.014501
124. M. Samanta, K. Pal, U.V. Waghmare, K. Biswas, Intrinsically Low Thermal Conductivity and High Carrier Mobility in Dual Topological Quantum Material, n-type BiTe. (2020). https://doi.org/10.1002/anie.202000343
125. J.R. Christman, Fundamentals of Solid State Physics. Wiley, New York, 1988.
126. A. Majumdar, P. Reddy, Role of electron–phonon coupling in thermal conductance of metal–nonmetal interfaces. Applied Physics Letters 84 (2004) 4768. https://doi.org/10.1063/1.1758301
127. D.T. Morelli, G.A. Slack, (2006). High Lattice Thermal Conductivity Solids. In: S.L. Shindé, J.S. Goela, (eds) High Thermal Conductivity Materials. Springer, New York, NY. https://doi.org/10.1007/0-387-25100-6_2
128. C.L. Julian, Theory of heat conduction in rare-gas crystals. Phys. Rev. 137 (1965) A128. https://doi.org/10.1103/PhysRev.137.A128
129. M.D. Nielsen, V. Ozolins, J.P. Heremans, Lone pair electrons minimize lattice thermal conductivity. Energy Environ. Sci. 6 (2013) 570. DOI: 10.1039/c2ee23391f
130. M.E. Fine, L.D. Brown, H.L. Marcus, Elastic constants versus melting temperature in metals. Scr. Metall. 18 (1984) 95. https://doi.org/10.1016/0036-9748(84)90267-9
131. J.S. Dugdale, D.K.C. Macdonald, Lattice Thermal Conductivity. Phys. Rev. 98 (1955) 1751. https://doi.org/10.1103/PhysRev.98.1751
132. J.K. Nrskov, T. Bligaard, J. Rossmeisl, C.H. Christensen, Towards the computational design of solid catalysts. Nature Chemistry 1 (2009) 37. https://doi.org/10.1038/nchem.121
133. G. Hautier, C. Fischer, A. Jain, T. Mueller, G. Ceder, Finding nature's missing ternary oxide compounds using machine learning and density functional theory. Chemistry of Materials 22 (2010) 3762. https://doi.org/10.1021/cm100795d

134. M. Calandra, A.N. Kolmogorov, S. Curtarolo, Search for high $T_c$ in layered structures: The case of LiB. Phys. Rev. B 75 (2007) 144506. https://doi.org/10.1103/PhysRevB.75.144506
135. W. Al-Sawai, H. Lin, R.S. Markiewicz, L.A. Wray, Y. Xia, S.Y. Xu, M.Z. Hasan, A. Bansil, Topological electronic structure in half-Heusler topological insulators. Phys. Rev. B 82 (2010) 125208. https://doi.org/10.1103/PhysRevB.82.125208
136. L. Sun, Y. Gao, B. Xiao, Y. Li, G. Wang, Anisotropic elastic and thermal properties of titanium borides by first-principles calculations. J. Alloys Compd. 579 (2013) 457. https://doi.org/10.1016/j.jallcom.2013.06.119
137. S. Kerdsongpanya, B. Alling, P. Eklund, Effect of point defects on the electronic density of states of ScN studied by first-principles calculations and implications for thermoelectric properties. Phys. Rev. B 86 (2012) 195140. https://doi.org/10.1103/PhysRevB.86.195140
138. K.H. Bennemann, J.W. Garland, in Superconductivity in *d*- and *f*- Band Metals, edited by D.H. Douglas, AIP Conf. Proc. No. 4, Edited by D. H. Douglass (AIP, New York, 1972), p.103.
139. K.H. Lee, K.J. Chang, First-principles calculations of the Coulomb pseudopotential $\mu^*$: Application to Al. Phys. Rev. B 52 (1995) 1425. https://doi.org/10.1103/PhysRevB.52.1425
140. N.E. Christensen, D.L. Novikov, Calculated superconductive properties of Li and Na under pressure. Phys. Rev. B 73 (2006) 224508. https://doi.org/10.1103/PhysRevB.73.224508
141. P. Cudazzo, G. Profeta, A. Sanna, A. Floris, A. Continenza, S. Massidda, E.K.U. Gross, *Ab Initio* Description of High-Temperature Superconductivity in Dense Molecular Hydrogen. Phys. Rev. Lett. 100 (2008) 257001. https://doi.org/10.1103/PhysRevLett.100.257001
142. M. Fox, Optical properties of solids. New York: Academic Press; 1972.
143. M.A. Hadi, S.H. Naqib, S.R.G. Christopoulos, A. Chroneos, A.K.M.A. Islam, Mechanical behavior, bonding nature and defect processes of $Mo_2ScAlC_2$: a new ordered MAX phase. J. Alloys. Compd. 724 (2017) 1167. https://doi.org/10.1016/j.jallcom.2017.07.110
144. F.L. Hirshfeld, Bonded-atom fragments for describing molecular charge densities. Theoretica Chimica Acta 44 (1977) 129. https://doi.org/10.1007/BF00549096
145. R.D. Harcourt, Diatomic antibonding σ* s orbitals as ′metallic orbitals□ for electron conduction in alkali metals. J. Phys. B 7 (1974) L41. DOI: 10.1088/0022-3700/7/2/003
146. V.K. Srivastava, Ionic and covalent energy gaps of CsCl crystals. Physics Letters 102A (1984) 127. https://doi.org/10.1016/0375-9601(84)90795-3
147. J.C. Phillips, Ionicity of the chemical bond in crystals. Rev. Mod. Phys. 42, 317 (1970). https://doi.org/10.1103/RevModPhys.42.317
148. C. Wang, S. Liu, H. Jeon, J.-H. Cho, Origin of charge density wave in the layered kagome metal $CsV_3Sb_5$. Phys. Rev. B 105 (2022) 045135. https://doi.org/10.1103/PhysRevB.105.045135
149. C. Guo, C. Putzke, S. Konyzheva, X. Huang, M. Gutierrez-Amigo, I. Errea, D. Chen, M.G. Vergniory, C. Felser, M.H. Fischer, T. Neupert, P.J.W. Moll, Switchable chiral transport in

charge-ordered kagome metal $CsV_3Sb_5$. Nature volume 611 (2022) 461. https://doi.org/10.1038/s41586-022-05127-9

150. Z. Liang, X. Hou, F. Zhang, W. Ma, P. Wu, Z. Zhang, F. Yu, J.-J. Ying, K. Jiang, L. Shan, Z. Wang, X.-H. Chen, Three-Dimensional Charge Density Wave and Surface-Dependent Vortex-Core States in a Kagome Superconductor $CsV_3Sb_5$. Phys. Rev. X 11 (2021) 031026. https://doi.org/10.1103/PhysRevX.11.031026

151. C. Wang, S. Liu, H. Jeon, Y. Jia, J.-H. Cho, Charge density wave and superconductivity in the kagome metal $CsV_3Sb_5$ around a pressure-induced quantum critical point. Phys. Rev. Materials 6 (2022) 094801. https://doi.org/10.1103/PhysRevMaterials.6.094801